\documentclass[authoryear,preprint,11pt]{elsarticle}
\usepackage{amssymb}
\usepackage{eurosym}
\usepackage{color,colortbl}
\usepackage{bm}
\usepackage{booktabs}
\usepackage{multirow}
\usepackage{longtable}
\usepackage{arydshln}
\usepackage{comment}

\usepackage{makecell}
\usepackage{enumitem}
\usepackage{subcaption}
\usepackage{graphicx}
\usepackage[a4paper, total={8in, 8in}]{geometry}
\usepackage{subcaption}
\usepackage{tikz}
\usepackage{pgfplots}
\usetikzlibrary{arrows,calc}
\usepackage{graphicx}
\usepackage{fancybox}
\usepackage{pifont}
\usepackage{siunitx}

\usepackage{multirow}
\usepackage{lscape}
\usepackage[round]{natbib}

\usepackage{amsthm,amsmath,amssymb,eurosym}

\newtheorem{definition}{Definition}

\journal{European Journal of Operational Research}

\usepackage{lineno, blindtext}

\begin{document}

\begin{frontmatter}
     \title{A multiple criteria methodology for prioritizing \\ and selecting portfolios of urban projects}
    \author[osm,corl]{{\sc M. Barbati}\corref{cor1}}\ead{maria.barbati@port.ac.uk}
     \author[ist]{{\sc J.R. Figueira}}
    \author[deb,corl,osm]{{\sc S. Greco}}
      \author[osm,corl]{{\sc A. Ishizaka}}
     \author[osm]{{\sc S. Panaro}}
    \address[deb]{Department of Economics and Business, University of Catania, Sicily, Italy}
    \address[ist]{CEG-IST, Instituto Superior T\'{e}cnico,  Universidade de Lisboa, Portugal}
    \address[corl]{ Centre of Operations Research and Logistics (CORL), \\ University of Portsmouth,  Portsmouth, United Kingdom}
    \address[osm]{Faculy of Business and Law, \\ University of Portsmouth,  Portsmouth, United Kingdom}
    \cortext[cor1]{Corresponding author at: Faculty of Business and Law, University of Portsmouth, Richmond Building, Portland Street, PO13DE, Porsmouth, UK.}

    \begin{abstract}
    \noindent  This paper presents an integrated methodology supporting  decisions in urban planning.  In particular, it deals with the prioritization and the selection of a portfolio of projects related to buildings of some values for the  cultural heritage in cities. In particular, our methodology has been validated to the historical center of Naples, Italy. Each project is evaluated on the basis of a set of both quantitative and qualitative criteria with the purpose to determine their level of priority for further selection. This step was performed through the application of the {\sc{Electre Tri-nC}} method. This method is a multiple criteria outranking based model for ordinal classification (or sorting) problems and allows to assign a priority level to each project as an analytical ``recommendation'' tool.  A set of resources (namely budgetary constraints) as well as some logical constraints related to urban policy requirements have to be taken into consideration together with the priority of projects in a portfolio analysis model permitting to identify the efficient portfolios and to support the selection of the most adequate set of  projects to activate. The  process has been conducted  thanks to the interaction between analysts, municipality representative and specialists. The proposed methodology is generic enough to  be applied in other territorial or urban planning problems.
    \end{abstract}
    \vspace{0.25cm}
\end{frontmatter}

\vfill\newpage

\section{Introduction}\label{sec:introduction}
\noindent Today the cities have to face big challenges due to the contemporary environmental, socioeconomic and institutional crisis \citep{garcia2017cultural}, that add up to the already problematic context \citep{rees1996urban}. Indeed, in the last century the urban sprawl has grown impetuously determining some difficulties for urban sustainability. Often the natural resources have been so much eroded  that is not possible to divide the city from the rural or natural areas.  The urban space has been expanded without providing adequate services (e.g., public transportation or waste collection) and infrastructures (e.g., social housing or leisure spaces) \citep{mcgreevy2017complexity}. In this context, the objectives and the constraints of the urban policies such as the approach to define them have to be consequently revised.

With respect to the \emph{objectives}, the general paradigm for urban policies is changing, taking into account the need to control the further geographic expansion of the cities and the use of more systemic and holistic approach in planning \citep{hansen2015uptake}. The attention is  moved, consequently, to urban sustainability focusing on:
\begin{itemize}[label={--}]
    \item Management and optimization of resource  \citep{agudelo2012harvesting}.
    \item Waste and pollution reduction, liveability improvement \citep{newman1999sustainability}.
    \item Enhancement of natural and cultural heritage \citep{gravzulevivciute2006cultural}.
\end{itemize}

With respect to the \emph{constraints}, a definition of urban policies have also to consider some relevant restrictions that are becoming more restrictive and selective. Among these, the current reduction of public expenditure to implement the planned policies has a specific relevance, which calls for the definition of a proper prioritization of actions to be funded.

In this paper we propose a general methodology to define urban policies that takes into account the above remarks and, therefore, is characterized by:
\begin{enumerate}
    \item A formulation in terms of a multiple criteria problem allowing the consideration of a family of heterogeneous standpointss.
    \item Consideration of integrated families of actions, technically called portfolios \citep{SaloEtAl11}, as feasible solutions of the problem at hand.
    \item An interactive procedure for the optimization problem as well as in the sorting problem in order to involve stakeholders, specialists and policy makers in the whole decision process.
    \item Integration in the optimization problem of a sorting procedure to assign a priority level to each feasible action.
\end{enumerate}

Despite its  general applicability, the above methodology was applied to a real world resources optimization for  the regeneration of a large and complex  historical centre in the city of Naples (Italy). Therefore, we will present below the real and specific arguments that helped us to identify the four methodology features listed above. In Naples, a large part of the historical centre has been inscribed as World Heritage Site, with the objective to safeguard the identity of communities worldwide. At the same time, other objectives, related to the preservation of the cultural heritage, have to be pursued. These objectives can be both with:
\begin{itemize}[label={--}]
    \item Economic nature, such as promotion of tourism flows \citep[e.g.,][]{MckercherEtAl05}, and local entrepreneurships and business \citep[e.g.,][]{TuanNa08}; and,
    \item Non-economic nature \citep{Blake00}, such as  social inclusion \citep{VasileEtAl15} or community engagement \citep{Waterton15}.
\end{itemize}
	
\noindent This suggests to adopt a multicriteria optimization procedure to choose among feasible actions, that is point $1)$ in the list of characteristics of our methodology.

Moreover, in  view of an integrated vision, feasible actions should not be considered as isolated but as parts of a comprehensive and unique  program to be evaluated in its totality. The output of the decision will be the definition of a portfolio of actions to implement. This suggests to approach the problem at hand in terms of portfolio decision analysis, that is point $2)$ in the list of characteristics of our methodology.

UNESCO requires the shared understanding of cultural heritage, which requires that, not only specialists and policy makers, but also stakeholders and, more in general, citizens, should be involved in its process of enhancement and protection.  This suggests to adopt an interactive procedure for the decision problem, that is point $3)$ in the list of characteristics of our methodology.

For all sites inscribed in the World Heritage List, UNESCO requires a management plan with  the identification and allocation of necessary resources. However, optimizing and collecting the  resources necessary for the management plan of sites that include a complex and big area as a historic centre is often quite problematic. In case of Naples, the problem is exacerbated because  the local authorities (often the principal owners) do not have enough resources for preserving huge cultural heritage. This requires an accurate, specific, and selective prioritization of actions in order to distinguish those that are deferrable from those ones requiring a prompt intervention. This suggests to define these priorities with a well established sorting procedure that has to be combined in the selection of the optimal portfolio of action, that is point $4)$ in the list of characteristics for our methodology.  In fact, this represents the most innovative contribution in the methodology we are proposing.

In this sense emerges the possibility of using Multiple Criteria Decision Aiding (MCDA) methodologies \citep[see e.g.,][]{greco2016multiple,ishizaka2013multi} that can guide the Decision-Makers (DMs) throughout the process. In the literature several methodologies have been introduced to deal with real cases of preservation of the cultural heritage. The main scope is the identification and the selection of the possible actions to be made. Usually, several criteria are considered  as in \cite{WangZe10} where the cultural aspect was taken into account together with economic, architectural, environmental, social and continuity aspects.

Among the others, Analytic Hierarchic Process (AHP) was adopted by \cite{WangZe10} for the selection of the reuse of historic buildings in Tapei City. Again, in \cite{HongCh17}, AHP  helps to rank the historic buildings to be reused with a different scope in the Grand Canal Area in China. Also, \cite{KututEtAl14} used AHP combined with the additive ratio assessment method for the definition of preservation actions for the historic buildings in Vilnius or in \cite{lolli2017} where is combined with the $K$-means algorithm for selecting the energy requalification interventions.

Often, the weighted-sum of the criteria is considered as in \cite{DuttaHu09} for the selection of historic buildings to preserve in the historical city of Calcutta or in \cite{GiulianiEtAl18} to classify the possible reuse of historic grain silos in Italy. In the same perspective, \cite{FerrettiEtAl14} used the multi-attribute value theory for ranking the different reuse of mills located in the metropolitan city of Turin.

Furthermore, \cite{GioveEtAl11} tested some MCDA methods, as the use of the Choquet integral, to evaluate the sustainability of  projects for the reuse of the Old Arsenal in Venice.

Some authors also proposed to integrate MCDA methodologies within a GIS environment as in \cite{TarraguelEtAl12} or \cite{GirardTo07} or with an integrated spatial assessment as in \cite{cerreta2010}. \cite{HamadoucheEtAl14}  integrated them  to select the sites that needs urgent conservation in an archaeological site in Algeria. Similarly, \cite{OppioEtAl15} integrated spatial MCDA analysis with GIS and SWOT analysis for studying the reuse of castles in Valle D'Aosta region.
Another aspect to consider is the inclusion in the decision process of several stakeholders as in \cite{YungCh13} for the preservation of historic buildings in Hong Kong or in \cite{cerreta2017perceived} for the  design of resilient landscapes.

While all these works are referring to single case studies with only some insights provided on how to extend the methodologies to different applications, \cite{FerrettiCo15} have proposed an integrated framework involving MCDA methods for the evaluation of heritage systems. They stress the importance of the interaction with the specialists and the DMs, in a transparent process, and how MCDA must support public authorities in the definition of their strategic planning.

In this sense, \cite{NesticoEtAl17} models the investment selection on historic buildings as a knapsack problem. The adoption of such an approach allows to define a plan that does not consider only one projects but the whole portfolio of projects. This strand of methodologies, called portfolio decision analysis \citep{SaloEtAl11}, helps DMs to make more informed decision and it has been implemented in several contexts from location of wind farms \citep{CranmerEtAl18} to research and development project selection \citep[e.g.,][]{CauglarGu17} and it can be integrated with many other multiobjective methods as in \cite{barbati2018}.

We aim to integrate a portfolio decision problem with an MCDA sorting methodology and in particular with the {\sc{Electre Tri-nC}} \citep{AlmeidaDiasEtAl12}, embedded in a transparent and interactive process, to allow the assessment and the prioritization of the investment in cultural heritage.
We aim to show how the methodology can help DMs to rationalize their decisions. We use it in a cultural leading city in Europe, Naples (Italy), and in particular to its historical center where a huge number of interventions has been planned but with the lack of assigning priority to the interventions. We show how our approach works also through the interaction with different actors.

The paper is organized as follows. In Section \ref{sec:context}, we illustrate the context of the historical center of Naples city, while in Section \ref{sec:methodology} we illustrate our methodology. In Section  \ref{sec:casestudy}, we list the stakeholders, the actors involved, the criteria and the projects of the case study.  In Section \ref{sec:Interaction}, we explain how the interaction with the different actors was conducted while in Section \ref{sec:selectingport}, we formulate the portfolio selection problem. In Section \ref{sec:experiments}, we report the results of our experiments. Finally, in Section \ref{sec:insights}, we discuss some insights coming from the practice and we conclude the paper  .

\section{Context}\label{sec:context}
\noindent The historic city center of Naples has been inscribed as World Heritage Site in 1995. Naples is a major city in the southern of Italy and it is among the most ancients cities in Europe. Its structure and its culture have been created throughout the history with influences from several civilizations, from its Greek foundation in 470 B.C., through the Roman period, to the Aragonese period until the 19th century. A multitude  of governmental and ecclesiastic buildings are still well preserved. The ancient city center is well recognizable with its Greek rectangular grid layout and its Aragonese walls. Moreover, the underground is made of several cavities used in the course of the years for different aims. Furthermore, we find peculiar attractions and many businesses activities.

In addition to this part of the cultural heritage there is a long tradition of intangible heritage. In fact, Naples has been for long periods the leading cultural city attracting artists and scholars in philosophy, art, literature, music, and theater. Its influence it is still well-known worldwide and events such as exhibits and shows are planned with continuity.  Recently, the image of the city has been relaunched in Italy and worldwide, attracting an increasing number of tourists.

Being a World UNESCO site, the city of Naples approved a management plan for its historic  center which  defines the vision, the general objectives and the approach  to preserve the universal values  associated to the cultural heritage \citep[see][]{comune2011}. This plan highlights the stakeholders involved, the strategic challenges and  lists a series of possible improvements of the area concerning both tangible and intangible cultural heritage. Despite the plan indicates a  set of projects to implement and their associated benefits and disadvantages, it fails to indicate which projects should be prioritized and which are the most beneficial to implement the strategic challenges previously mentioned. Anyway, from the plan it does not emerge a whole picture that would help to decide which investments are the most fundamental ones and, on the contrary, it is assumed that all the projects will be implemented. Unfortunately, in a period in which the rationalization of the expenses and resources is an increasing need, the realistic view is that not all the projects will be funded or will be funded in different periods. Furthermore, in the long time passed since the development of the plan and the obtaining of the funds, the number of projects has grown and had to be revised.
 For example, in the last years, some projects planned have been detailed and others have been added. Based on the management plan, funds are being searched by the local authorities.
Indeed, a systematic methodology could help the DMs to rationalize their choice, when the funds become available.

In particular, this seems a case in which we need to select a portfolio of projects in a complex process in which more than a single criterion needs to be considered for their  prioritization and subsequently their selection. A systematic procedure that could support the DMs in this process should be employed to make informed decisions, integrating economic and non economic aspects. This is even truer in the case of cultural heritage where the economic aspect of the problem needs to be integrated with criteria related to very different aspects often not tangible and difficult to quantify. A multiple criteria methodology helps the DMs throughout the whole process supporting his decision even in presence of qualitative evaluations.

\section{An integrated methodology}\label{sec:methodology}
\noindent Our methodology is based on the integration of a portfolio decision problem with a sorting procedure. In particular, the methodology includes the following steps:

\begin{enumerate}
    \item \emph{Description of the urban planning problem}. From the analysis of the urban context  the following elements should be identified:
    \begin{itemize}[label={--}]
        \item The participants in the decision process and their specific aims.
        \item The actions to implement in the urban context, i.e. the projects, that could be implemented.
        \item The criteria, i.e. the standards used to judge the projects.
        \item The evaluation of every single project for each criterion.
        \item The constraints on the projects (i.e., budget constraints).
    \end{itemize}

    \item \emph{Prioritization}. Using a sorting method, the projects will be prioritized according to the criteria defined. In particular, we suggest to use a multicriteria sorting method in a constructive perspective \citep{AlmeidaDiasEtAl10}, in which the stakeholder representatives are involved in the decision model building process. In this way, we assign to each project a priority, that is used in the subsequent portfolio optimization problem.
    \item \emph{Selection of a portfolio of projects}. We build an optimization problem in which we maximize the number of projects  with the highest priorities, taking into account some constraints related to the specific characteristics of the decision problem at hand (e.g., budget constraints). In this way, we define  ``the best portfolio". This best portfolio should be presented and recommended to the participants as their best option in the current environment.
    \item \emph{Robustness Analysis}. We should perform an analysis for what concerns the variations of the parameters of the model \citep{roy1998,roy2010robustness}. Indeed, we aim to verify that even if we change some of the parameters used in the model, e.g., the weights representing the importance of the criteria or the formulation of some of the constraints, the participants will still be comfortable with the solution proposed.

\end{enumerate}

In the following we report, for our case study, a description of each step including how the interaction with the actors involved (specialists, analysts, a municipality representative, and a specific specialist that played the role of DM) has been considered .

\begin{enumerate}
	\item  \textit{Description of the urban planning problem}. The first  preliminary activity has been the identification of the participants. We clarify what they are expecting to obtain from the implementation of the regeneration works for the tangible cultural heritage projects. This step has been implemented thanks to a careful analysis of the management plan published by the municipality.  Then, we have first identified the criteria for our analysis from the management plan and, later, we have  particularized them thanks to the interaction with the specialists.  The projects and their evaluations were detailed in relation to the  available information.
    \item \textit{Prioritization}. Thanks to a socio-technical approach based on the interaction with different actors and the implementation of the  {\sc{Electre Tri-nC}} method \citep{AlmeidaDiasEtAl12}, we define for each project its own category, prioritizing the most important ones according to the set of criteria defined. This step, therefore, is composed of several stages:
    \begin{itemize}[label={--}]
        \item Definition of the criteria weights through the interaction with the specialists.
        \item Construction of the reference actions and the categories to which assign the projects thanks to the interaction with the DM and the analysts.
        \item Modeling the imperfect knowledge of data and arbitrariness through the construction of the discriminating (indifference and preference) thresholds thanks to the interaction with the DM  and the analysts.
        \item Definition of the veto trough the interaction with DM and analysts.
       \item Application of the {\sc{Electre Tri-nC}} ordinal classification method.
    \end{itemize}
     \item \textit{Modelling the selection of projects}. We have defined a binary linear programming model that includes a specific objective function that optimizes the number of projects with highest priority and a a set of specific constraints derived from the interaction with the Municipality representative and the DM. The application of this model determines the portfolio of projects to implement. The results have been discussed again with the DM.
        \item \textit{Performing  a robustness analysis}. We have tried several scenarios, changing criteria weights and veto thresholds, that we have discussed with the DM and the municipality representative.
\end{enumerate}

\noindent In the following we can describe the application of our methodology for the case study.

\section{A case study in the city area of Naples}\label{sec:casestudy}
\noindent In this section, we present the concrete elements of our case study: the participants, the criteria and the set of projects related to the urban tangible cultural heritage to be evaluated.

\subsection{Participants listed in the management plan of the historic city centre: Their aims and expectations}\label{sec:Participants}
\noindent Several participants are involved in the decision aiding process and they have different interests generated by the re-qualification of the historic city center of Naples. The various participants can be listed in three main categories: institutional bodies, social and cultural organizations, and local businesses. A brief description for each of them is reported below.

\begin{enumerate}
  \item Institutional bodies (which include  Naples Municipality, Naples Province, Campania Region, Ministry of Cultural Heritage, the Port authority and local healthcare authority): Their aim is the conservation and enhancement of the cultural heritage (tangible and intangible), the improvement of the local economy and wellbeing for the citizens. The governmental bodies often own the buildings of which they want to improve the use. Therefore, they are promoters of the interventions and have active interest in managing  the buildings. Moreover, they expect that the whole local and international community will benefit from the general improvement of the area.
  \item Social and cultural organizations (which include, for example, Universities, ecclesiastic communities, and nonprofit organizations): Their aim is to reduce the number of degraded/abandoned buildings, to promote the local culture and the traditional knowledge. They want to increase the cultural offer and the spaces for  activities such as exhibitions, shows and local community events. They expect also that the reuse of the buildings (of which sometimes are owners)  increases number of services for the citizens, students and visitors, contributing to the liveability and social cohesion of the area.
  \item Local businesses (which include, for example, building companies, tourism and hospitality companies, craft associations): Their aim is to promote the restoration of the buildings and to provide new services to satisfy the increasing flow of tourists. They want to improve the general aspect of the historical city center making it a better tourist attraction safeguarding the traditional artisanship and local products. They are also interested in the general improvement of the city. They expect that the increasing number of visitors will bring additional income thanks to the opening of new activities or the improvement of traditional ones, such as artisans shops.
\end{enumerate}

\subsection{Criteria identified in our study}\label{sec:Criteria}
\noindent Criteria were built to make operational the different standpoints that were identified from the values and concerns of the participants. These were adapted from the management plan for the historical center of Naples city, in Italy.
The management plan for the historical center of Naples city has identified four significant standpoints (or strategic challenges) for the conservation and the enhancement of the UNESCO site \citep[see][]{comune2011}. In our study, we have taken into account  adjusted different standpoints, to make operational the  adapted criteria, always according to the values and concerns of the participants to our specific context.

There are four major standpoints in our study: (1) preservation of tangible cultural heritage; (2) promotion of the traditional craftsmanship, tourism, and local businesses; (3) improvement of the quality of the urban environment; and (4) social benefits for community. These standpoints along with their respective criteria (and possible subcriteria) are presented next.

\begin{enumerate}
	\item Preservation of tangible cultural heritage standpoint.  \\
    In this standpoint participants are concerned with the preservation of tangible cultural heritage.  These aspects can be represented by the following two criteria.
    	\begin{enumerate}
      		\item Compatibility of the project (notation: $g_1$; label: {\tt CON-COMP}; unit: qualitative levels; preference direction: to maximize). This criterion consists of an evaluation of the compatibility of  the establishment with the project defined. For each project we consider the compatibility with the proposed use  for the establishment. This criterion can be represented through a 4-level qualitative scale with the following verbal assertion and respective labels: very high (VH), high (H), medium (M), low (L).
     		\item Growing the usability of the cultural heritage (notation: $g_2$; label: {\tt CON-USAB}; unit: percentage; preference direction: maximize). This criterion comprises the increasing in terms of usability for the building. We use a percentage scale in order to consider t the fact that beyond a certain level a project cannot enhance any more the usability of the heritage.
        \end{enumerate}
    \item Promoting traditional craftsmanship-Tourism-Business standpoint. \\
    In this standpoint participants consideres the behavior that should be taken for nurturing traditional craftsmanship assests and the growing of the production of local products, as well as tourism and the establishment of new local businesses. This is important to improve the local economy, preserving the intangible heritage and  the local traditions.
    	\begin{enumerate}
        	\item Promotion of the traditional craftsmanship and traditional knowledge (notation: $g_3$; label: {\tt PRO-CRAF}; unit: qualitative levels; preference direction: to maximize). This criterion considers the enrichment in the historical city center of the traditional craftsmanship and knowledge, and also the enhancement of the offer of local products. It includes two subcriteria:
            	\begin{enumerate}
            		\item  New jobs development and
                         increasing of local
                         products. Use of a 4-level qualitative scale as in criterion $1.a$.             .
                    \item Dissemination of                       intangible cultural
                           heritage. Availability of projects to promote traditional knowledge.  The same type of scale as in the previous subcriterion is adopted.
            	\end{enumerate}
   The two subcriteria can be merged to create the scale of the criterion.  The first subcriterion is more significant than the second one. By merging the two sub-criteria scales, we can have a 16-level qualitative scale for the criterion: VH-VH(16), VH-H(15),...,H-VH(12),...,M-VH(8),...,L-VH(4),...,L-L(1). The numbers from 16 to 1 are only used to code the profiles so that, e.g.,   16 is not a value, but a code for VH-VH. Indeed, taking into account all profiles resulting form the 16 combinations of ordinal evaluations on the two criteria, we can define only a partial ordering for which combination ($\alpha$,$\beta$) is dominating combination ($\gamma$,$\delta$) if $\alpha$ is not worse than $\gamma$ and $\beta$ is not worse than $\delta$. In Figure \ref{Fig:Hasse} we represent the $16$ possible  profiles  with respect to the two considered subcriteria on a Hasse diagram where it is possible to see all the possible preference relations among the different possible profiles. More precisely. the profile $(\alpha,\beta)$ is preferred to the profile $(\gamma,\delta)$ if there is an arrow directed from  $(\alpha,\beta)$ to $(\gamma,\delta)$ (as it is the case, e.g. for 16 and 12) or if there is a set of consecutive arrows starting from $(\alpha,\beta)$ and arriving to  $(\gamma,\delta)$  (as it is the case, e.g., for $12$ and $6$ which are linked through the arrows going from $12$ to $11$, from $11$ to $7$, and from $7$ to $6$). Observe also that, if it is not possible to reach  $(\gamma,\delta)$ starting from $(\alpha,\beta)$ with one or more consecutive arrows, then $(\alpha,\beta)$ and $(\gamma,\delta)$ are not comparable (as it is the case, e.g. for $14$ and $4$).

\begin{figure}\label{Fig:Hasse}
 \centering
\includegraphics[width=11.5cm]{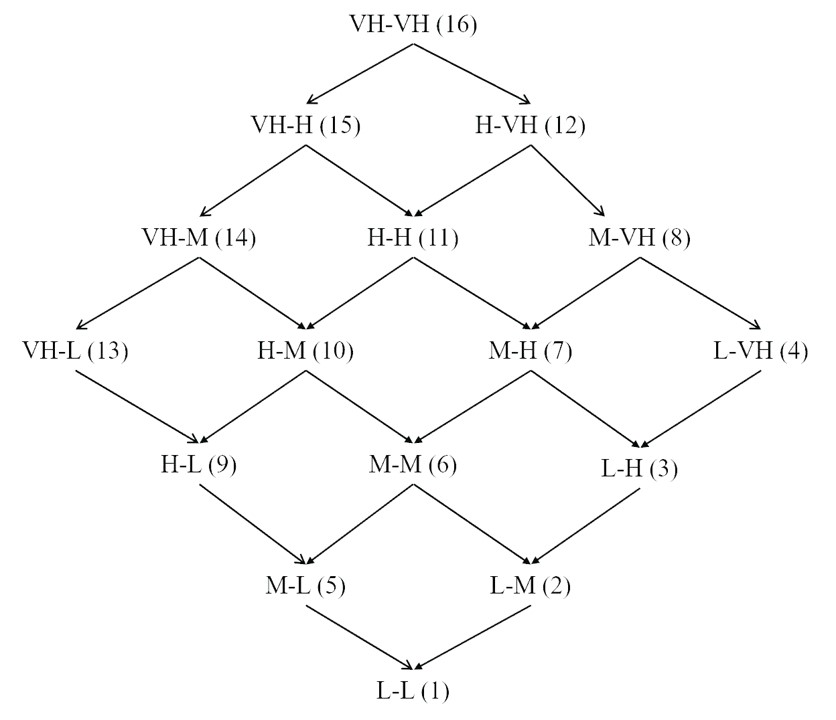}
 \caption{Hasse diagram profiles for the two subcriteria of criterion $g_3$}
 \label{Fig:Hasse}
\end{figure}

            \item Promotion of local entrepreneurship and businesses (notation: $g_4$; label: {\tt PRO-BUSI}; unit: qualitative levels; preference direction: maximize). New local companies created, use of a 4-level qualitative scale as in criterion $1.a$.
            \item Promotion of tourism (notation: $g_5$; label: {\tt PRO-TOUR}; unit: qualitative levels; preference direction: to maximize). Flow of visits in the city. Use of a 4-level qualitative scale as in criterion $1.a$.
        \end{enumerate}
    \item Urban environment standpoint. \\
    In this standpoint participants relate to the urban aesthetics as a significant factor to improve the perception of Historical Urban Landscape and its preservation. In this  case it provides the assessment of projects as refurbished facades, surrounding urban spaces and green spaces. There is only one criterion to make operational the quality of urban environmental standpoint.
    	\begin{enumerate}
        	\item Maintenance of urban spaces (notation: $g_6$; label: {\tt ENV-MAIN}; unit: $m^2$; preference direction: to maximize). This criterion consists of the urban area that has been restyled.
        \end{enumerate}
    \item Social benefits for community standpoint. \\
    In this standpoint, participants are concerned with improving the identity of local communities and their knowledge on cultural heritage. This is a standpoint of utmost importance and it is rendered operational using two different criteria as stated below.
    	\begin{enumerate}
        	\item Enrichment of the cultural offer (notation: $g_7$; label: {\tt SOC-CULT}; unit: qualitative levels; preference direction: to maximize). This criterion consists of initiatives to enrich the  cultural offer, aiming to increase the public awareness and cultural identity of the city.
            	\begin{enumerate}
            		\item New cultural proposals
                          target. Use of a 4-level qualitative scale as in criterion $1.a$
                    \item Increasing of entrances at museums and other cultural sites. Use of a 4-level qualitative scale as in criterion $1.a$.
            	\end{enumerate}
The overall scale of this criterion is built as in $2.a$.
            \item Social cohesion (notation: $g_8$; label: {\tt SOC-COHE}; unit: quantitative; preference direction: maximize). Number of non-profit organizations participating in management of tangible cultural heritage and their planned cultural activities. Use of a 4-level qualitative scale as in criterion $1.a$
        \end{enumerate}
\end{enumerate}

\subsection{Actors involved in our study: Municipality representative and specialists}\label{sec:Representatives}
\noindent In this subsection we present the participants involved in our study (specialists, analysts, and a municipality representative), the interaction with them and the contribution of this interaction.
\noindent As in the studies developed by \cite{BotteroEtAl15} and \cite{BotteroEtAl18}, a very interesting situation for dealing with this type of problems is when we can interact with all the participants views and work within a focus group scheme. In the current situation having a representative from each stakeholder was not possible. For this reason, we have interacted with different actors, as follows:

\begin{enumerate}

\item First, we have interviewed an specialist for each of the standpoints presented in the previous subsection. We have asked to rank criteria for each of them as we explain in Section \ref{sec:weights}.

\item Second, we have asked these four specialists to interact in a focus group scheme. In this way, we have gathered the different standpoints that are essential to consider when dealing with such a difficult problem. The specialists, that suggested in the first phase different rankings for the criteria have found an agreement and they have very well accepted the methodology.

\item Third, we have dialogued with a local municipality representative that has a through knowledge on  the  ongoing projects in the historic city center. He has provided relevant information about the details and the peculiarities of the projects. He has also highlighted difficulties and restrictions linked to the different projects.

\item Finally, we have interacted with a different specialist with more technical knowledge about the problem, the projects and the data. This specialist has contributed throughout all the decision process with a continuous interaction with the analysts. For this reason he has played the role of the DM in this study, while the authors played the role of analysts.
\end{enumerate}

Therefore, the interaction with the actors had two significant steps:
\begin{itemize}[label={--}]
    \item interaction between the analysts, the specialists and the municipality representative which has permitted to clarify the elements of the real problem and collect information about preferences and constraints;
    \item  interaction  between the analysts and the specialist playing the role of DM which has permitted to model the problem and to select a possible portfolio of projects.
\end{itemize}

\subsection{Tangible urban cultural heritage: Projects}\label{sec:Actions}
\noindent In this subsection we listed for the selected projects the typology of the tangible cultural heritage, e.g., a building and area and so on. Moreover, we think it is appropriate to highlight what is the state of the building, e.g., usable, degraded or , and what type of intervention has been planned, e.g., improving use, making it usable, or establishing new functions.

\begin{enumerate}
   \item \textit{Murazione Aragonese di Porta Capuana} (notation: $a_{1}$; label: {\tt Mura-Capua}; Typology: Historic Entrance; State:  Degraded; Intervention: Improving use.). Recovery of the wall surrounding the historic marble door  named ``Porta Capuana", including the new pedestrians areas.
\item \textit{Castel Capuano} (notation: $a_{2}$; label: {\tt Cast-Capua}; Typology: Castle; State:  Partially usable; Intervention: Partial reuse). Recovery of lower ground floor and ground floor to open a museum of human rights and laws.
\item \textit{Complesso ex-ospedale di Santa Maria della Pace} (notation: $a_{3}$; label: {\tt Comp-Maria}; Typology: Monumental complex; State:  Partially usable; Intervention: Opening and reuse). Recovery of the historic building for the creation of a center for cultural/leisure activities, along with elderly and students accommodation.
\item \textit{Complesso S. Lorenzo Maggiore} (notation: $a_{4}$; label: {\tt Comp-Loren}; Typology: Monumental complex; State:  partially usable; Intervention: Improving use).
Recovery of the historical archive of the municipality of Naples and some parts of the complex to create a cultural centre including youth training activities.
\item \textit{Complesso S. Gregorio Armeno ed ex Asilo Filangieri} (notation: $a_{5}$; label: {\tt Comp-Grego}; Typology: Monumental complex; State:  Partially usable; Intervention: Improving use). Enhancement of the archaeological area and allocate  some parts of the complex for the cultural/leisure activities.
\item \textit{Insula del Duomo - Area archeologica} (notation: $a_{6}$; label: {\tt Area-Duomo}; Typology: Archaeological site; State:  Degraded; Intervention: Opening).
Opening the archaeological area to visitors.
\item \textit{Complesso di S. Lorenzo maggiore (Area archeologica)} (notation: $a_{7}$; label: {\tt Area-Loren}; Typology: Archaeological Site; State:  Partially usable; Intervention: Improving use). Increasing the archaeological area which can be visited.
\item \textit{Teatro antico di Neapolis} (notation: $a_{8}$; label: {\tt Teat-Neapo}; Typology: Archaeological site; State:  Archaeological excavation yard; Intervention: New use).
Restoration of the ancient Roman theater to make it open to the public.
\item \textit{Chiesa SS. Cosma e Damiano} (notation: $a_{9}$; label: {\tt Chie-Cosma}; Typology: Church; State:  Degraded; Intervention: New use). Recovery of the historical church for the creation of a centre that improves the social and leisure activities of the neighborhood.
\item \textit{Castel dell'Ovo} (notation: $a_{10}$; label: {\tt Cast-D'Ovo}; Typology: Castle; State:  Partially usable; Intervention: Improving use).
Restoration of a part of the castle for improving use and sightseeing tours.
\item \textit{Complesso dei Girolamini} (notation: $a_{11}$; label: {\tt Comp-Gerol}; Typology: Monumental complex; State:  Degraded; Intervention: Improving use). Enhancement of the monumental site.
\item \textit{San Gioacchino a Pontenuovo} (notation: $a_{12}$; label: {\tt SanG-Ponte}; Typology: Historical building; State:  Usable; Intervention: Improving use). Improving of the historical archive and promotion of the center for cultural events.
\item \textit{Sant'Aniello a Caponapoli} (notation: $a_{13}$; label: {\tt SanA-Capon}; Typology: Church; State:  Usable; Intervention: Improving use). Creation of a UNESCO  documentation centre.
\item \textit{Complesso Trinit\`a' delle Monache} (notation: $a_{14}$; label: {\tt Comp-Monac}; Typology: Monumental complex; State:  Degraded; Intervention: New use). Restoration and recovery of the historical site for creation of multi-purpose center.
\item \textit{Mercatino S.Anna Di Palazzo} (notation: $a_{15}$; label: {\tt Merc-Palaz}; Typology: Market; State:  Degraded; Intervention: Opening and reuse). Creation of services for local companies and youth center.
\item \textit{Chiesa San Giovanni Battista delle Monache} (notation: $a_{16}$; label: {\tt Chie-Monac}; Typology: Church; State:  Degraded; Intervention: Improving use). Restoration of the church for the creation of the ``Aula Magna" for the Academy of the Fine Arts.
\item \textit{Complesso Santa Maria della Fede} (notation: $a_{17}$; label: {\tt Comp-Mfede}; Typology: Monumental complex; State:  Degraded; Intervention: Improving use). Recovery of the complex for the creation of a center for cultural/leisure activities, and also students accommodation.
\item \textit{Carminiello al Mercato} (notation: $a_{18}$; label: {\tt Carm-Merca}; Typology: Monumental complex; State:  Partially usable; Intervention: Improving use). Creation of a services center for the local commercial activities and businesses.
\item \textit{Complesso di S. Paolo Maggiore} (notation: $a_{19}$; label: {\tt Comp-Paolo}; Typology: Monumental complex; State:  Partially usable; Intervention: Improving use).
Restoration of the cloister and intervention on the degraded part of the complex for creation of a museum and spaces for cultural/social activities.
\item \textit{Villa Ebe alle rampe di Lamont Young} (notation: $a_{20}$; label: {\tt Vill-EbeRa}; Typology: Villa; State:  Degraded; Intervention: Opening and reuse).
Restoration of an abandoned villa for the creation of the tourist center.
\end{enumerate}

\begin{figure}\label{fig:Centre-Napoli}
 \centering
\includegraphics[height=13.36cm,width=16.24cm]{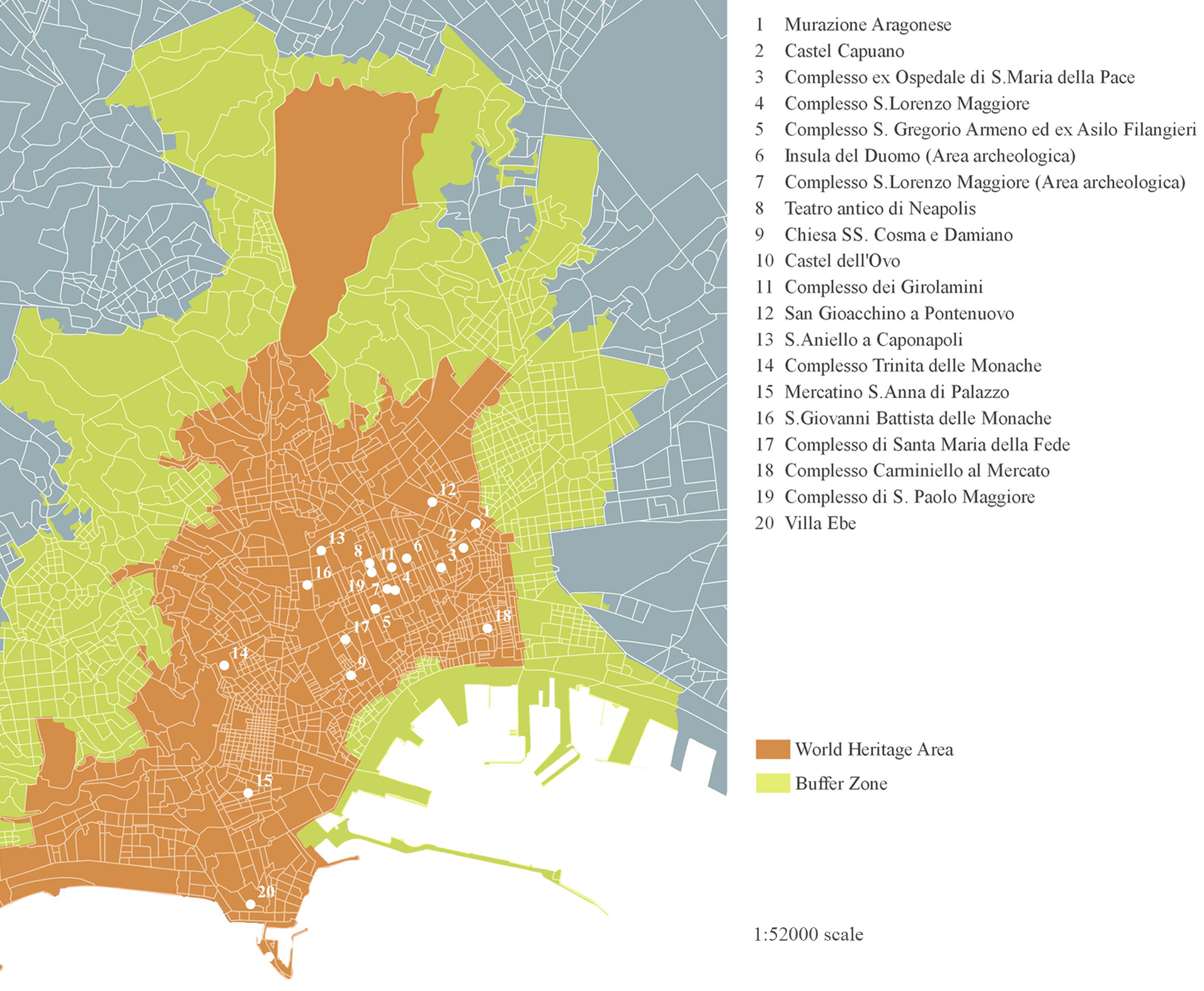}
 \caption{Projects and their locations}
 \label{fig:Centre-Napoli-Sites}
\end{figure}

\noindent In Figure \ref{fig:Centre-Napoli-Sites} we can see the location of all the sites in the city map of Naples.

Moreover, for each of the project one or more functions, intended as the planned activities, have been identified. These are classified as:

\begin{itemize}[label={--}]
\item Tourist facilities (label $U_1$);
\item Museum and archaeological sites (label $U_2$);
\item Students and elderly accommodation (label $U_3$);
\item Leisure Activities (label $U_4$);
\item Record Offices Archives (label $U_5$);
\item Citizenship facilities (label $U_6$).

\end{itemize}

\noindent The functions have been listed for each project  in Table \ref{tab:Functions} in Appendix \ref{sec:Appendix_Data}.

\subsection{The performance table}\label{sec:table}
\noindent To evaluate the performance of  projects we  analysed them in detail. Some information was not directly available; therefore, we have estimated it. For example, the quantitative criterion related to the maintenance of the urban spaces has been judged considering the area of the interventions and the surrounding areas that will benefit from it.  Please see  Appendix \ref{sec:Appendix_Data} for Table \ref{tab:PerformanceTabl} with all the evaluations for each project and each criterion.


\section{Interaction with actors involved: A behavioral and socio-technical approach}\label{sec:Interaction}
\noindent In this Section, we present a method to assign priority levels to the tangible urban heritage projects. These projects have different levels of priority to be implemented and they are not obvious to assign since they are based on several conflicting criteria.

\subsection{Why an \sc{Electre Method} ?}\label{sec:why-electre}
\noindent Before choosing an MCDA method, we checked for some basic requirements that eventually led to the choice of an {\sc{Electre}} method \citep[for more details see][]{FigueiraEtAl13}. The requirements are mainly the following:

\begin{itemize}[label={--}]
	\item The number and the nature of criteria. In our problem we have eight criteria, which fits perfectly within the adequate number for using an MCDA method (in general, in between five and twelve). In addition, in the problem there are both quantitative and qualitative criteria. The scales are rather heterogeneous with different units: $m^2$, $K$\euro, Number, and verbal levels.
    \item The choice of  the scales for some criteria is rather arbitrary, as for example for the criterion {\tt PRO-CRAF} . In addition, it not easy to define the concrete meaning of all the qualitative levels of some scales. Thus, some imperfect knowledge of data related with arbitrariness when building the criteria as imprecision or uncertainty is present.
       \item Some examples with dummy buildings, e.g., buildings that do not exists with evaluation or characteristics  defined  only for the aim of the analysis,  were presented to the DM to test his sensitivity to compensatory phenomena and we conclude there was a  non relevance of compensatory effects.
    \item When comparing two different buildings, the DM pointed out the advantages of one building (reasons for) with respect to the second as well as the disadvantages (reasons against) of the first with respect to the second over.
    \item Through the use of some dummy projects, e.g.,  fictitious projects that do not exist and have ad-hoc built characteristics in terms of implementation and restoration of cultural heritage,  we could observe that the DM could prefer a project over a second one, the second one over a third, but the third could be preferred over the first. It means that intransitivities are accepted.
\end{itemize}

All these aspects are characterizing our problem, for this reason, as summarized in \cite{FigueiraEtAl16}, the choice of an {\sc{Electre}} method is the most appropriate one among the variuos MCDA methods.

Since the problem is by its very nature a sorting problem, the next question was, which one of the {\sc{Electre}}  methods for sorting problem was the most adequate. We started by asking one of the specialists if it was possible to delimit the categories by (lower and upper) limiting profiles as in {\sc{Electre Tri-B}} and {\sc{nB}} \citep[see][]{FernandezEtAl17}. This was rather difficult for the specialist. Then, we tried to see if a more or less ``central'' reference action for each category could be easier to identify  as in  {\sc{Electre Tri-B}} \citep[see][]{AlmeidaDiasEtAl10}. Again, the specialist did not feel comfortable and was not able to define a ``central'' action. When we gave him the possibility of choosing any representative reference action per category, he felt much more comfortable to provide such information, which led us to adopt the {\sc{Electre Tri-nC}}  \citep{AlmeidaDiasEtAl10} and pursue our study with this method. A brief description of the method has been provided in Appendix \ref{Appendix:Electre-nC}.

\subsection{Determining the discriminating (indifference and preference) thresholds}\label{sec:weights}
\noindent For  the indifference and preference thresholds, also called discriminating thresholds, we adopt the procedure proposed in \cite{RoyEtAl14}. These thresholds are used for modeling the imperfect knowledge of data and should result from the interaction between the analysts and the specialist.

\subsection{Constructing the reference actions}\label{sec:reference-actions}
\noindent The reference actions were defined in a co-constructive way with the participation of the specialist and the analysts. A graphical representation helped in making an adequate choice of the representative alternatives of each category. The categories of priority with respect to conservation, promotion, environmental and social standpoints, are the following.

\begin{itemize}[label={--}]
	\item Category $C_1$:  This category contains the projects with low priority.
    \item Category $C_2$:  This category contains the projects with medium priority.
    \item Category $C_3$: This category contains the projects with high priority.
    \item Category $C_4$: This category contains the projects with very high priority.
\end{itemize}

First, the analysts started by asking the DM if he could identify in the performance table representative projects of each categories, or at least for some of the categories. This was not possible. He found the task difficult given the set of projects provided in the performance table.

Then, along with the DM, the analysts
made a graphical representation of all criteria scales, one after the other, in the same sheet of paper and asked if he could draw a representative project of category $C_1$. After some reflection they draw an action with the following profile $b_{1,1} = [L(1), \, 20\%,\, M-L(5),\, L(1),\, L(1),\, 1000m^2,\, M-L(5),\, L(1)]$.

The process was repeated for the other categories. Table \ref{tab:refact} presents the set of representative actions that were  co-constructed during the interaction process between the analysts, from one side, and the DM, from the other side.

\begin{table}[htb!]
\centering
 \begin{tabular}{ccccccccc} \hline\hline
  Representative  & $g_1$ & $g_2$ & $g_3$ & $g_4$ &
  $g_5$ & $g_6$ & $g_7$ & $g_8$  \\
  Actions   & {\scriptsize {\tt CON-COMP}} & {\scriptsize {\tt CON-USAB}} & {\scriptsize {\tt PRO-CRAF}} & {\scriptsize {\tt PRO-BUSI}} & {\scriptsize {\tt PRO-TOUR}} & {\scriptsize {\tt ENV-MAIN}} &  {\scriptsize {\tt SOC-CULT}} & {\scriptsize {\tt SOC-COHE}}  \\ \hline
  $b_{1,1}$ & 1 & 20 & 5  & 1 & 1 & 1000  & 5  & 1\\ \hdashline
  $b_{2,1}$ & 2 & 40 & 7  & 2 & 1 & 3000& 7  & 2\\
  $b_{2,2}$ & 2 & 30 & 6  & 2 & 1 & 3500& 6  & 3\\ \hdashline
  $b_{3,1}$ & 3 & 70 & 11 & 3 & 2 & 10000& 11 & 4\\
  $b_{3,2}$ & 3 & 50 & 7 & 2 & 2 & 5000& 11 & 3\\ \hdashline
  $b_{4,1}$ & 4 & 80 & 12 & 3 & 3 & 30000& 15 & 4\\ \hline\hline
 \end{tabular}
 \caption{Reference actions}\label{tab:refact}
\end{table}

\subsection{Determining the veto thresholds}\label{sec:veto}
\noindent The process for assessing the veto is similar to the one used for the discriminating thresholds. The concepts were clearly explained to the specialist playing the role of DM in an illustrative example and were well accepted. The veto thresholds were easily assessed obtaining the values presented in Table \ref{tab:veto} in Appendix \ref{sec:Appendix_Data}. It was, of course, more laborious than the task for assessing the variable thresholds. Let us remark that for some criteria,  no discriminating threshold has been identified. Veto threshold along with the weights of criteria and preference based parameters  are related with the role each criterion plays in the construction of the outranking relation.

\subsection{Determining the weights of criteria}\label{sec:weights}
\noindent Criteria have different importance. The procedure applied for determining the relative importance of criteria is the Simos-Roy-Figuiera (SRF) method proposed in \cite{FigueiraRo02}. The interaction was conducted first with each specialist and after in the focus group.

In this way, we have the opportunity to perform a series of analyses considering even more than one set of weights, defining different versions of the problem for which the solution will be acceptable, and therefore robust, even having different set of values for the weights.

In our case, in the focus group, the specialists provided the following information for each step, respectively:
\begin{enumerate}
\item The provided ranking  is the following where the symbol $\prec$ is used to denote ``strictly less important than'', and $\sim$ to denote ``equally important than'' :
    	\[
        	{\displaystyle g_5 \prec \{g_2 \sim g_4\}
            \prec \{g_3 \sim g_7\} \prec \{g_6 \sim g_8\}
            \prec g_1}
        \]
\item After, a rich discussion and many exchanges, a  decision was taken on the following distribution of blank cards that are displayed between brackets:
    \[
        	{\displaystyle g_5 \; [1] \; \{g_2 \sim
            g_4\} \; [1] \; \{g_3 \sim  g_7 \} \; [0] \;
            \{g_6 \sim g_8 \} \;  [0] \; g_1}
    \]
\item Finally, we asked the specialists to let us know how many times the most important criterion, $g_1$, is more important than the least important one, $g_5$. This was a rather difficult question. We rephrased in the following way. If we assign a vote to criterion $g_5$ how many votes do you assign to criterion $g_1$? The specialists, after some reflections, say 10. This answered our first question.

\end{enumerate}

\noindent This classification leads to find the weights denoted by $w_{j}^{1}$ and reported in Table \ref{tab:weightstable}.

Anyway, after some comments and a quite long discussion, the specialist provided also an additional ranking and classification. In particular, they felt that they could change the position in the ranking of some criteria as follows:

\[
        	{\displaystyle g_5 \prec \{g_2 \sim g_7\}
            \prec g_4 \prec g_3 \prec \{g_6 \sim g_8\}  \prec \ g_1 }
        \]

They also said that in this case they will not add any blank cards and the distance between the first and the last level should be equal to 10. This different classification of criteria lead to a different set of weight called $w_{j}^{2}$ and reported in Table \ref{tab:weightstable}.

In Table \ref{tab:weightstable} we also have added the weights originated by the individual interaction with  every specialist, in particular:
\begin{itemize}[label={--}]
\item Specialist in preservation of tangible cultural heritage standpoint (label:{\tt{ EXP-CONV}}), weights $w_{j}^{3}$;
\item Specialist in promotion of the traditional craftsmanship and local products standpoint (label:{\tt{ EXP-PROM}}), weights $w_{j}^{4}$;
\item Specialist in  quality of urban environment standpoint (label:{\tt{ EXP-URBE}}), weights $w_{j}^{5}$;
\item Specialist in social benefits for community standpoint (label:{\tt{ EXP-COMM}}), weights $w_{j}^{6}$;
\end{itemize}

 The ranking of criteria that have generated those weights has been reported from Table \ref{tab:expconv} to Table \ref{tab:expcomm} in Appendix \ref{sec:Appendix_Data}.

\begin{table}[htb!]
\centering
 \begin{tabular}{ccccccccc} \hline\hline
  Criteria  & $g_1$ & $g_2$ & $g_3$ & $g_4$ &
  $g_5$ & $g_6$ & $g_7$ & $g_8$  \\
  weights   & {\scriptsize {\tt CON-COMP}} & {\scriptsize {\tt CON-USAB}} & {\scriptsize {\tt PRO-CRAF}} & {\scriptsize {\tt PRO-BUSI}} & {\scriptsize {\tt PRO-TOUR}} & {\scriptsize {\tt ENV-MAIN}} &  {\scriptsize {\tt SOC-CULT}} & {\scriptsize {\tt SOC-COHE}}  \\ \hline
  $w_{j}^{1}$    & 20.0 & 8.0 & 14.0 & 8.0 & 2.0 & 17.0  & 14.0 & 17.0  \\
  $w_{j}^{2}$    &22.7& 6.4& 14.5& 10.5& 2.3& 18.6& 6.4& 18.6   \\
  $w_{j}^{3}$    & 13.4 &	18.3&	6.1&	18.3&	6.1&	18.3&	13.4	&6.1
  \\
  $w_{j}^{4}$     & 11.3&	11.3	&16.1&	16.1&	1.6	&11.3&	16.1	&16.1
   \\
  $w_{j}^{5}$    &  4.3 &	14.7&	11.2&	11.2&	11.2&	21.5&	7.8	&18.1
  \\
  $w_{j}^{6}$    & 20.8	&6.3&	16.6&	4.2&	18.7&	12.5&	16.7&6.1
  \\ \hline\hline
 \end{tabular}
 \caption{Sets of criteria weights}\label{tab:weightstable}
\end{table}

\section{Selecting a portfolio  of reusable physical urban cultural heritage artifacts}\label{sec:selectingport}
\noindent  In this Section we describe how the selection of the highest priority projects is modelled. First, we introduced a particularized objective function, that maximize the number of projects to introduce in our portfolio with the highest priority. Second, we discussed with the municipality representative and the DM about the potential constraints that  should be considered for the definition of the urban planning.

\subsection{Objective function}\label{sec:obj-functions}
\noindent After assigning a priority level to each project we must construct the portfolio of projects proposed to be  funded. This decision problem was handled by defining a    $0-1$ knapsack problem with additional logical constraints related to budget limitations and urban planning requirements.  Since we cannot select all the projects at the same time due to the multiple constraints, we should start by selecting as much projects of possible with the highest priority, then  the ones with the second highest and so on. More precisely, we can associate a $0-1$ decision variable, $x_i$, to each $a_i \in A$, so that $x_i=1$ if $a_i$ is selected and $x_i=0$ otherwise.. Then, the number of projects in the maximal priority category $C_q$ are maximised,  solving the following optimization problem:

\[
	{\displaystyle \max f_q(x) = \sum_{\{i\; :
    \; a_i \in C_q\}}x_i},
\]

subject to all the constrains of the problem. Assume that the optimal value of this problem is $f^{\ast}_q = k_q$. Then, for the maximization of the number of artifacts in $C_{q-1}$, the second highest priority category, we can add the constraint $\sum_{\{i\; : \; a_i \in C_q\}}x_i=k_q$ to the initial set of constraints and proceed with the next optimization:

\[
	{\displaystyle \max f_{q-1}(x) = \sum_{\{i\; : \; a_i
    \in C_{q-1}\}}x_i}.
\]

\vspace{1cm}
\noindent The process is repeated until the lowest category is explored.

This sequential process can, however, be replaced by  the resolution of an equivalent single optimization problem. Instead of several objective functions we define a single objective  function as follows.

\[
	{\displaystyle \max f(x) = \sum_{\{i\; : \; a_i
    \in A\}}c_ix_i},
\]

\noindent where

\[
	c_i = \left\{
	\begin{array}{lcl}
    	\bar{c}_h = 1 & \mbox{for} & h = 1,\\
        & & \\
        {\displaystyle \bar{c}_h = 1 +
        \sum_{k=1}^{h-1}\bar{c}_k\vert C_k
        \vert} & \mbox{for} & h = 2,\ldots,q.\\
    \end{array}
    \right.
\]

We associate  a coefficient $\bar{c}_h$ with each category $C_h$. We define also the total weight of a category by multiplying the coefficient of the category by its number of elements, i.e., $\bar{c}_h\vert C_h\vert$.

The idea is that the value of the coefficient of category $C_h$ should be strictly larger the sum of all the total values $\overline{c}_k|C_k|$ associated to all the categories $C_k, k=1,\ldots,h-1,$ having a smaller priority.

With the proposed procedure, projects with highest priority are first selected, unless, it will not be possible due to the constraints. In such case, the optimization process goes to next priority level, until the lowest one.

\subsection{Constraints}\label{sec:constraints}
\noindent Constraints for the problem were discussed with the municipality representative and the specialist acting as DM. This allows to have a real perspective on the problem. While the municipality representative helped in understanding the broad problems of the area, the DM was able to provide a more detailed description of them, allowing the analysts to formulate the constraints. We have several constraints related to the different aspects.

The most important constraint is the one related to the budget available for the implementation of the projects. Each project $a_i \in A$ is associated with a cost $s_i$ (see Table \ref{tab:PerformanceTabl} in  Appendix \ref{sec:Appendix_Data} for the costs of each project). Given that at the moment the budget allocated to these  projects has still not be defined in a unique manner, the council representative and the DM have suggested to implement several scenarios with different available budget. Denoting  the available budget as $B$, this constraint can be formulated as:
\begin{equation}\label{con:costs}
\sum_{\{i:a_i \in A\}} s_ix_i\leqslant B.
\end{equation}

In addition, the DM and the municipality representative have highlightes that we can have constraints related to the type of functions that each building or area will have after its restoration/recovery. Furthermore a strategic element in the decision choice is represented by the location of the projects. On the basis of that, several logical constraints can be formulated.

As first, in the ancient city center there are three main roads called ``decumano", highlighted in Figure \ref{fig:Partizione}. The DM suggested to have a lower limit on the umber of regeneration projects to be activated in this part of the city.
From the map, we can see that the set of projects  $A_d=\{a_2, a_3, a_7, a_8, a_{11}$, $a_{19}\}$ are located on one of the``decumano". The DM suggested that among those projects a minimum amount of projects, let us say $Q_d$, should be carried out. We formulate this as:
\begin{equation}\label{con:decumano}
\sum_{\{i:a_i \in A_d\}} x_i\leqslant Q_d.
\end{equation}
The DM will be asked to supply a desired value for $Q_d$.

\begin{figure}[htp!]\label{fig:Partizione}
 \centering
\includegraphics[height=15.88cm,width=16.45cm]{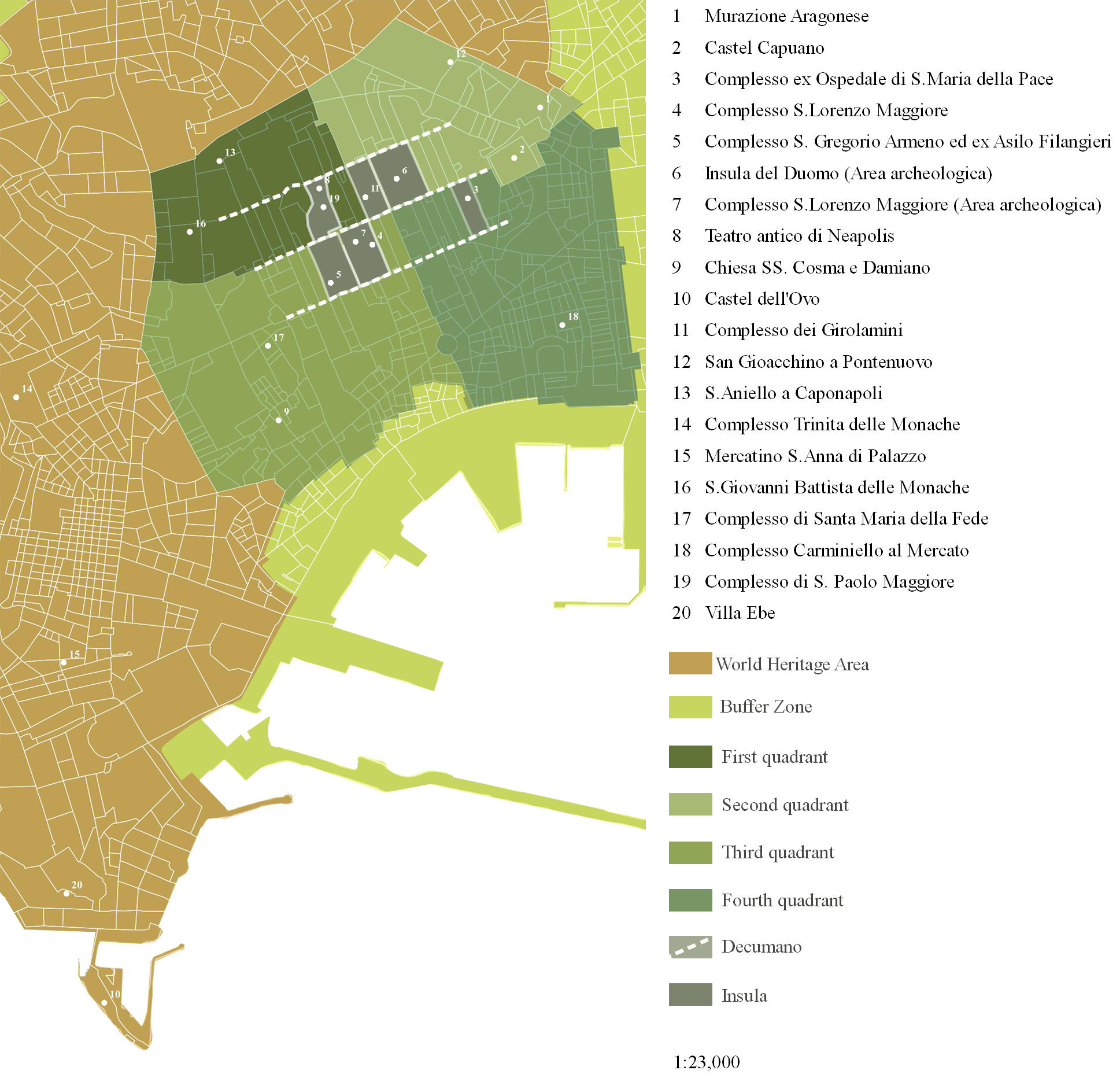}
 \caption{``Decumano'', ``Insula'' and Quadrants in the historic city center}
 \label{fig:Partizione}
\end{figure}

Second, the ancient city center can be divided in  smaller areas called ``insulae" derived from the intersection of the``decumano" with the smaller roads called ``cardines". In six of these areas are located some projects, this has been represented in Figure \ref{fig:Partizione}.
In our case, the DM has pointed out that if two projects are activated on the same ``insula" they can create a greater benefit  than the sum of their single criterion, generating a synergy, but he could not tell how much was the enhancement. For this reason, he said that he is going to take into account portfolios where at least a given number, called $N$, of those synergies will be adopted.  To model that, we define the set of ``insula" $I=\{I_1,\ldots,I_s, \ldots, I_6\}$ , such that $I_s \in A, s=1,\ldots,6$, contains all the projects in the $s$-th ``insula". and the $0-1$ decision variables $x_{ij}$  with $a_i \in I_s$ and $a_j \in I_s$  such that $x_{ij}=1$ if both $a_i$ and $a_j$ are activated and $x_{ij}=0$ otherwise. The following set of constraints, then, should be formulated:
\begin{equation}\label{con:siner1}
x_{ij}\geqslant x_i+x_j-1;
\end{equation}
\begin{equation}\label{con:siner2}
x_i\geqslant x_{ij};
\end{equation}
\begin{equation}\label{con:siner3}
x_j\geqslant x_{ij};
\end{equation}
\begin{equation}\label{con:siner4}
\sum_{\{i,j:a_i \in I_s, a_j \in I_s\}} x_{ij}\geqslant N, \hspace{1cm}\forall I_s \in I.
\end{equation}

Third, another important aspect are the functions that each building will have in relation to the area in which locates. For each project we have identified all the possible purposes  (see Table \ref{tab:Functions} in Appendix \ref{sec:Appendix_Data}). In addition, the DM  has also mentioned to consider bigger areas of the ``insula''  that represent four important zones of the ancient city center. Let us remark that each project can have one or more purposes. In particular, if we indicate with $K=\{1,\ldots, k, \ldots,6\}$ the set of the purposes, and with $U=\{1,\ldots,u,\ldots, 4\}$ the set of the four bigger areas, we can define the binary coefficients $z_{iuk}$  equals to 1 if the project $a_i$ is in bigger area $u$  delivering purpose $k$, $0$ otherwise.
The DM has indicated two types of constraints concerning those two aspects. For a specific purpose, a given number of projects $N_k$  should be implemented. This constraint can be modeled as:
\begin{equation}\label{con:function1}
\sum_{\{u \in U, a_i \in A\}} z_{iuk}x_i\geqslant N_k \hspace{1cm}\forall k \in K
\end{equation}

He has also suggested that it would be ideal to  distribute equally some of the purposes in some quadrants. For this purpose, we define the binary variables  $y_u$ $\forall u \in U$. The constraints are of the type:

\begin{equation}\label{con:function2}
	\sum_{\{i:a_i \in A:  z_{iuk}=1\}} x_i \geqslant 1-M y_u \hspace{1cm}\forall k \in K, \forall u \in U
\end{equation}
with $M$ being a big number, and
\begin{equation}\label{con:function3}
	\sum_{u} y_u \leqslant 4-q, q\in\{0,1,2,3\},
\end{equation}
with $q$ being a number included between $0$ and $3$.

Those two constraints imply that at least one project with a given purpose should be open in at least $4-q$ bigger area. For example, if $q$ is equal to 2, two bigger area should have at least one project implemented for one of the purposes defined.

The monobjective binary programming model can be solved with any linear optimization solver. We have used the CPLEX 12.1 software.

\section{Case study}\label{sec:experiments}
\noindent The results have been obtained thanks to the interaction with the DM throughout the whole decision process. First, we have carried out several possible scenarios for assigning priorities and, after that we have selected the projects to carry out.

\subsection{Assigning priority levels to the projects}\label{sec:prioritization}
\noindent According to the data reported in the Appendix \ref{sec:Appendix_Data}, we have used the MCDA-ULAVAL software \footnote{Available at http://cersvr1.fsa.ulaval.ca/mcda/.} that implements the majority of the {\sc{Electre }} methods, including the {\sc{Electre Tri-nC}} method allowing to insert all the data and parameters elicited.

First, the analysts suggested to apply the methodology for each set of weights from Table \ref{tab:weightstable} using the other  necessary data of  Appendix \ref{sec:Appendix_Data}. We have called the configuration $F_{w_{j}^{1}}$  to specify that we used weights $w_{j}^{1}$, $F_{w_{j}^{2}}$  for weights $w_{j}^{2}$ and so on. The results are reported in Table \ref{tab:category1}.

\begin{table}[h!]
\centering
 \begin{tabular}{ccccccc} \hline\hline
  Weights  & $w_{j}^{1}$ & $w_{j}^{2}$ & $w_{j}^{3}$ & $w_{j}^{4}$ &
  $w_{j}^{5}$ & $w_{j}^{6}$   \\
  Configuration   & $F_{w_{j}^{1}}$ & $F_{w_{j}^{2}}$ & $F_{w_{j}^{3}}$ & $F_{w_{j}^{4}}$ & $F_{w_{j}^{5}}$ & $F_{w_{j}^{6}}$   \\ \hline
$a_{1}$({ \tiny{\tt Mura-Capua}})&$C_3$&$C_3$&$C_3$&$C_3$&$C_3$&$C_3$\\
$a_{2}$({ \tiny{\tt Cast-Capua}})&$C_3$&$C_3$&$C_3$&$C_3$&$C_3$&$C_3$\\
$a_{3}$({ \tiny{\tt Comp-Maria}})&$[C_3, C_4]$&$[C_3, C_4]$&$[C_3, C_4]$&$[C_3, C_4]$&$[C_3, C_4]$&$[C_3, C_4]$\\
$a_{4}$({ \tiny{\tt Comp-Loren}})&$C_3$&$C_2$&$C_2$&$C_2$&$C_2$&$C_3$\\
$a_{5}$({ \tiny{\tt Comp-Grego}})&$C_3$&$C_3$&$C_3$&$C_3$&$C_3$&$C_3$\\
$a_{6}$({ \tiny{\tt Area-Duomo}})&$C_1$&$C_1$&$C_1$&$C_1$&$C_1$&$[C_1, C_2]$\\
$a_{7}$({ \tiny{\tt Area-Loren}})&$C_2$&$[C_1, C_2]$&$C_2$&$[C_1, C_2]$&$C_1$&$C_2$\\
$a_{8}$({ \tiny{\tt Teat-Neapo}})&$[C_3, C_4]$&$[C_3, C_4]$&$[C_3, C_4]$&$[C_3, C_4]$&$[C_3, C_4]$&$[C_3, C_4]$\\
$a_{9}$({ \tiny{\tt Chie-Cosma}})&$C_2$&$C_2$&$C_2$&$C_2$&$C_2$&$C_2$\\
$a_{10}$({ \tiny{\tt Cast-D'Ovo}})&$C_3$&$C_3$&$C_3$&$C_3$&$C_3$&$C_3$\\
$a_{11}$({ \tiny{\tt Comp-Gerol}})&$C_4$&$C_4$&$C_4$&$C_4$&$C_4$&$C_4$\\
$a_{12}$({ \tiny{\tt SanG-Corvo}})&$C_2$&$C_2$&$C_2$&$C_2$&$C_2$&$C_2$\\
$a_{13}$({ \tiny{\tt SanA-Capon}})&$C_2$&$[C_1, C_2]$&$C_2$&$C_2$&$[C_1, C_2]$&$[C_2, C_3]$\\
$a_{14}$({ \tiny{\tt Comp-Monac}})&$C_3$&$C_3$&$C_3$&$C_3$&$C_3$&$C_3$\\
$a_{15}$({ \tiny{\tt Merc-Palaz}})&$C_2$&$C_2$&$[C_2, C_3]$&$[C_2, C_3]$&$[C_2, C_3]$&$C_2$\\
$a_{16}$({ \tiny{\tt Chie-Monac}})&$C_2$&$C_2$&$C_2$&$C_2$&$C_2$&$[C_2, C_3]$\\
$a_{17}$({ \tiny{\tt Comp-Mfede}})&$C_3$&$C_3$&$C_3$&$C_3$&$C_3$&$C_3$\\
$a_{18}$({ \tiny{\tt Carm-Merca}})&$C_2$&$C_2$&$C_2$&$C_2$&$C_2$&$C_2$\\
$a_{19}$({ \tiny{\tt Comp-Paolo}})&$C_3$&$C_3$&$C_3$&$C_3$&$C_3$&$C_3$\\
$a_{20}$({ \tiny{\tt Vill-EbeRa}})&$C_3$&$C_3$&$C_3$&$C_3$&$C_3$&$C_3$\\
  \\ \hline\hline
 \end{tabular}
 \caption{Categorization for each defined set of weights  }\label{tab:category1}
\end{table}

At this point, the analysts showed these categorizations to the DM. He has identified that the most interesting classification, closer to his perspective, is provided by configuration $F_{w_{j}^{1}}$. It represents the closest one to his opinion in terms of projects that should be prioritized.

The analysts wanted to conduct some further analysis to validate the performance of this categorization and wished to perform some sensitivity analysis, as follows:
\begin{enumerate}

\item First, he has tried to increase the number of cards introduced in the ranking that has generated the weights of the preferred classification. In particular he has tried to increase the distance between the first,  the second and the third level adding five cards to every level. This scenario,  was analyzed because, during the focus group, there were many exchanges about the number of cards to insert between each level. Some of the specialists in the focus group highlighted how criteria in the first and second level should be the most important, but they did not provide enough arguments to convince the others to insert some additional cards.  Anyway, even if this change gets to different weights, it did not change the categorization of the projects. Moreover, he has attempted to put 20 cards, distributing them between the last two criteria. This time the categorization was equal to the one obtained with weights $w_{j}^{2}$  , i.e. configuration $F_{w_{j}^{2}}$.

\item Second, the analysts felt that it was worth testing if the ratio between the last and the first level was influencing the solution. He has increased this value to 20. Again a categorization identical to the preferred one was obtained.

\item Third, the analysts felt that in order to test the robustness of the categorization, the preference information and the veto thresholds should be changed. The analyst, performed the analysis inserting some different  preference information, veto thresholds and the reference actions are reported in the Appendix \ref{sec:datisecondi}. They used the weights  $w_{j}^{1}$ derived from the focus group (Configuration $F_{w_{j}^{1}}^P$) and the most different ones expressed by the specialist {\tt{EXP-URBE}} (Configuration $F_{w_{j}^{5}}^P$).
\end{enumerate}

\begin{table}[h!]\
\renewcommand*{\arraystretch}{1.3}
\centering
 \begin{tabular}{ccc} \hline\hline
  Weights  & $w_{j}^{1}$ & $w_{j}^{5}$    \\
   \renewcommand*{\arraystretch}{1.0}
 Configuration   & $F_{w_{j}^{1}}^P$ & $F_{w_{j}^{1}}^P$   \\ \hline
$a_{1}$({ \tiny{\tt Mura-Capua}})&$[C_3, C_4]$&$[C_3, C_4]$\\
$a_{2}$({ \tiny{\tt Cast-Capua}})&$[C_2, C_3]$&$C_2$\\
$a_{3}$({ \tiny{\tt Comp-Maria}})&$[C_3, C_4]$&$[C_3, C_4]$\\
$a_{4}$({ \tiny{\tt Comp-Loren}})&$[C_1, C_2]$&$C_1$\\
$a_{5}$({ \tiny{\tt Comp-Grego}})&$C_2$&$C_2$\\
$a_{6}$({ \tiny{\tt Area-Duomo}})&$[C_1, C_2]$&$[C_1, C_2]$\\
$a_{7}$({ \tiny{\tt Area-Loren}})&$[C_1, C_2]$&$[C_1, C_2]$\\
$a_{8}$({ \tiny{\tt Teat-Neapo}})&$C_3$&$C_3$\\
$a_{9}$({ \tiny{\tt Chie-Cosma}})&$[C_3, C_4]$&$[C_3, C_4]$\\
$a_{10}$({ \tiny{\tt Cast-D'Ovo}})&$C_2$&$C_2$\\
$a_{11}$({ \tiny{\tt Comp-Gerol}})&$C_4$&$C_4$\\
$a_{12}$({ \tiny{\tt SanG-Corvo}})&$C_3$&$[C_2, C_3]$\\
$a_{13}$({ \tiny{\tt SanA-Capon}})&$[C_2, C_3]$&$[C_2, C_3]$\\
$a_{14}$({ \tiny{\tt Comp-Monac}})&$C_4$&$C_4$\\
$a_{15}$({ \tiny{\tt Merc-Palaz}})&$[C_2, C_4]$&$[C_2, C_4]$\\
$a_{16}$({ \tiny{\tt Chie-Monac}})&$[C_1, C_2]$&$[C_1, C_2]$\\
$a_{17}$({ \tiny{\tt Comp-Mfede}})&$C_3$&$[C_3, C_4]$\\
$a_{18}$({ \tiny{\tt Carm-Mercato}})&$C_2$&$C_2$\\
$a_{19}$({ \tiny{\tt Comp-Paolo}})&$C_3$&$C_3$\\
$a_{20}$({ \tiny{\tt Vill-EbeRa}})&$[C_2, C_4]$&$[C_2, C_4]$\\
  \\ \hline\hline
 \end{tabular}
 \caption{Categories obtained with a second set of veto thresholds, reference actions and preference information}\label{tab: categorization2}
\end{table}

These two new obtained categorizations were discussed with the DM. Both  configurations were considered not adequate. This is due to the presence of many projects that can be assigned to two different categories and in some cases  to even more categories (i.e., $a_{13}$ and $a_{15}$). The DM commented that this categorization does not represent what he retains being a good prioritization. In particular, concerns were raised for the presence of project $a_{14}$ in the last category. Even if it will restore the largest building and surrounding area (highest value for criterion $g_6$), project $a_{14}$ does not represent the DM's opinion of being a project to prioritize and it should be in a different category in comparison to project $a_{11}$. Also project $a_{10}$ has been classified in the category $C_2$, while the specialist felt that this project should be at least in the upper category being  a strategic attraction for the city. At this point the analyst felt that the best option was to move to the selection of the projects adopting the priorities obtained with the first set of reference of veto thresholds, reference actions and preference information.

\subsection{Selection}\label{sec:selection}
\noindent The analysts started with the preferred categorization, generated by configuration $F_{w_{j}^{1}}$ as this was indicated by the DM as the most preferred one.

First, the computation of the objective function was discussed.
Being the calculation of the coefficients $c_h$ depending on the number of categories, the analysts asked the DM if he wanted to assign the projects  between two categories  to one of the two categories. The DM did not feel that this operation was useful but he preferred to consider this intermediate categories as a single different category. Considering that, the calculation of the $c_h$ coefficients was carried out for every possible configuration (these are reported in Table \ref{Tab:chcoeff} in Appendix \ref{sec:Appendix_Model}).

After this, the DM and the analysts have considered which constraints should be included among the ones defined in Section \ref{sec:constraints} fixing the values of every parameter included in the constraints.

Starting from the budget constraint, the DM has expressed interest in conducting a set of simulations with different possible budgets available. This because, given the economic situation, he felt that one possibility was to verify what are the projects that will be selected considering those different scenarios. This allows  the DM to learn depending from the different amount of budget available what are the projects that will be selected on the basis of the assigned priorities. The analysts have suggested that the scenarios with the following budget could be tested. We indicate with $B(o)$ the possible budget available, and $F_{w_{j}^{1}}^{B(o)}$ the configuration generated with weights ${w_{j}^{1}}$ and Budget $B(o)$  (see Table \ref{tab:budget}).

\begin{table}[h!]\label{weightstable}
\centering
 \begin{tabular}{ccccccccc} \hline\hline \\
     Budget &$B(1)$ & $B(2)$ & $B(3)$ & $B(4)$ &
  $B(5)$ & $B(6)$ &B(7)
  \\ \hline
 & 52240	&45710	&39180	&32650	&26120	&19590	&13060
\\
  \\ \hline\hline
 \end{tabular}
 \caption{Budgets available  in $K$\euro }\label{tab:budget}
\end{table}

Next,  he specified that the ideal portfolio will have at least  four projects in the ``decumano", therefore in constraint (\ref{con:decumano}) the $Q_d$ is set equal to 4.

For the third set of constraints (\ref{con:siner1})-(\ref{con:siner4}) the DM specified that at least one synergy among the ones presented in every ``insula" should be verified. For this reason, parameter $N$ was set to 1 for constraint  (\ref{con:siner4}).

The DM has highlighted that, for what concerns the functions of the projects, he would like to have at least three projects for the function $U_3$, i.e., students and elderly accommodation, in constraint (\ref{con:function1})).

For the set of constraints (\ref{con:function2})-(\ref{con:function3}) the DM has indicated that he will be considering a good portfolio the one with at least one project that  provides facilities for the citizens and one project that provides facilities for the tourists in at least three of the four quadrants. This way parameter $q$ in constraint (\ref{con:function3}) should be set equal to 1.
The formulation of all these constraints is reported in the Appendix \ref{sec:Appendix_Model}.
The resolution of the binary programming model for the different available budgets has lead to the following portfolios in Table \ref{tab:selbudgetweights1} (Note that the symbol \checkmark means that a project should be carried out, while the symbol $\times$ denotes that the project should not be implemented).

Let us underline that for budgets $B(6)$ and $B(7)$ there is no possible solution to the model that satisfies at the same time all the constraints. Therefore, we have  asked the DM to remove or weaken some of the constraints. He has advised to lower the value of the projects expected to be in the``decumano" to 1 and the number of students and elderly housing facilities to 1. The portfolios reported in the Table \ref{tab:selbudgetweights1} have been obtained considering these two relaxed constraints.
As expected, the number or projects to be implemented decrease with the decreasing of the available budget. The DM was invited to reflect on the portfolios identified and he has provided the following comments:

\begin{table}[h!]
\centering
 \begin{tabular}{cccccccc} \hline\hline
  \\
  Configuration   & $F_{w_{j}^{1}}^{B(1)}$ & $F_{w_{j}^{1}}^{B(2)}$ & $F_{w_{j}^{1}}^{B(3)}$ & $F_{w_{j}^{1}}^{B(4)}$ & $F_{w_{j}^{1}}^{B(5)}$ & $F_{w_{j}^{1}}^{B(6)}$ & $F_{w_{j}^{1}}^{B(7)}$ \\ \hline
$a_{1}$({ \tiny{\tt Mura-Capua}})&\checkmark  & \checkmark & \checkmark & \checkmark & $\times$ & \checkmark & \checkmark\\
$a_{2}$({ \tiny{\tt Cast-Capua}})&\checkmark  & \checkmark & $\times$ & $\times$ & $\times$ & $\times$ & $\times$\\
$a_{3}$({ \tiny{\tt Comp-Maria}})&\checkmark  & \checkmark & \checkmark & \checkmark & \checkmark & $\times$ & $\times$\\
$a_{4}$({ \tiny{\tt Comp-Loren}})&\checkmark  & \checkmark & \checkmark & $\times$ & $\times$ & $\times$ & $\times$\\
$a_{5}$({ \tiny{\tt Comp-Grego}})&\checkmark  & \checkmark & \checkmark & \checkmark & \checkmark & \checkmark & \checkmark\\
$a_{6}$({ \tiny{\tt Area-Duomo}})&\checkmark  & $\times$ & $\times$ & $\times$ & $\times$ & $\times$ & $\times$\\
$a_{7}$({ \tiny{\tt Area-Loren}})&\checkmark  & $\times$ & $\times$ & $\times$ & \checkmark & $\times$ & \checkmark\\
$a_{8}$({ \tiny{\tt Teat-Neapo}})&\checkmark  & \checkmark & \checkmark & \checkmark & \checkmark & \checkmark & $\times$\\
$a_{9}$({ \tiny{\tt Chie-Cosma}})&\checkmark  & $\times$ & $\times$ & $\times$ & $\times$ & $\times$ & $\times$\\
$a_{10}$({ \tiny{\tt Cast-D'Ovo}})&\checkmark  & \checkmark & \checkmark & \checkmark & $\times$ & \checkmark & $\times$\\
$a_{11}$({ \tiny{\tt Comp-Gerol}})&\checkmark  & \checkmark & \checkmark & \checkmark & \checkmark & \checkmark & \checkmark\\
$a_{12}$({ \tiny{\tt SanG-Corvo}})&\checkmark  & \checkmark & \checkmark & \checkmark & \checkmark & \checkmark & \checkmark\\
$a_{13}$({ \tiny{\tt SanA-Capon}})&\checkmark  & \checkmark & $\times$ & $\times$ & $\times$ & \checkmark & \checkmark\\
$a_{14}$({ \tiny{\tt Comp-Monac}})&$\times$  & $\times$ & $\times$ & $\times$ & $\times$ & $\times$ & $\times$\\
$a_{15}$({ \tiny{\tt Merc-Palaz}})&\checkmark  & \checkmark & $\times$ & \checkmark & \checkmark & $\times$ & $\times$\\
$a_{16}$({ \tiny{\tt Chie-Monac}})&\checkmark  & \checkmark & \checkmark & \checkmark & $\times$ & $\times$ & $\times$\\
$a_{17}$({ \tiny{\tt Comp-Mfede}})&\checkmark  & \checkmark & \checkmark & \checkmark & \checkmark & $\times$ & $\times$\\
$a_{18}$({ \tiny{\tt Carm-Mercato}})&\checkmark  & $\times$ & $\times$ & $\times$ & $\times$ & $\times$ & $\times$\\
$a_{19}$({ \tiny{\tt Comp-Paolo}})&\checkmark  & \checkmark & \checkmark & \checkmark & $\times$ & $\times$ & $\times$\\
$a_{20}$({ \tiny{\tt Vill-EbeRa}})&\checkmark  & \checkmark & \checkmark & $\times$ & $\times$ & $\times$ & $\times$\\
  \\ \hline\hline
 \end{tabular}
 \caption{Projects selected for different budget available, weights $w_{j}^{1}$ }\label{tab:selbudgetweights1}
\end{table}

\begin{itemize}[label={--}]
\item The portfolio of configuration $F_{w_{j}^{1}}^{B(1)}$  was welcomed by the DM. Indeed, the DM  said that in a scenario with such available budget  he will prefer not to  carry out project $a_{14}$ but still implementing all the others, in order to improve most of the areas of the historic city center. In his standpoint, being project $a_{14}$ the one with the highest cost, it is not an extremely attractive project and he will happily compromise  in doing all the other projects.
\item The portfolio of configuration $F_{w_{j}^{1}}^{B(6)}$ could represent a good compromise because it is suggesting to implement all the most important projects with a reasonable budget that is more likely to be obtained.
\item The portfolio of configuration $F_{w_{j}^{1}}^{B(4)}$ has some areas of concerns due to the presence of projects $a_{12}$, $a_{15}$ and $a_{16}$ belonging to the  category $C_2$ and not including projects from the upper category $C_3$. Then, the analysts have solved again the model, excluding those projects from the optimization procedure. Anyway, once showed the different portfolio obtained  to the DM, he advised that the original portfolio was more representative of his preferred portfolio for that amount of budget available.
\item The DM considered as a good portfolios even the one for configuration $F_{w_{j}^{1}}^{B(5)}$. In fact, even with a very limited budget all the listed constraints will still be satisfied and the best projects of the last two categories will still be in the portfolio.
\item For the portfolios of configurations $F_{w_{j}^{1}}^{B(5)}$, $F_{w_{j}^{1}}^{B(6)}$ and  $F_{w_{j}^{1}}^B(7)$  the DM has disclosed some concerns about the quality of these portfolios with very few projects. He has also raised concerns on the fact that projects in lower categories were introduced. The analysts explained that this was due to the distribution of functions constraint (\ref{con:function2})-(\ref{con:function3}) and that he could try to do some other simulations relaxing also these constraints. Eventually, the DM really thought that this is a major and fundamental constraint that he will not compromise on happily, therefore he  accepted well the portfolio obtained.
\end{itemize}

The analysts asked the DM if he was satisfied with the analysis concerning the different budget scenarios and he replied that he had enough information that allow him to better understanding the process and to have a better grasp of how much funding will be necessary to have a portfolio that will be the mostly close to the preferred one.

\subsection{Robustness Analysis}\label{sec:Robustness}
\noindent A robustness analysis is important to verify that the results obtained  can be replied even in slightly changed conditions \citep{roy1998,roy2010two}. We focus on taking into account the different configurations derived by the different opinions of the specialists involved.

\begin{table}[htb!]
\centering
 \begin{tabular}{ccccccc} \hline\hline
  Configuration   & $F_{w_{j}^{1}}^{B(2)}$ & $F_{w_{j}^{2}}^{B(2)}$ & $F_{w_{j}^{3}}^{B(2)}$ & $F_{w_{j}^{4}}^{B(2)}$ & $F_{w_{j}^{5}}^{B(2)}$ & $F_{w_{j}^{6}}^{B(2)}$  \\ \hline
$a_{1}$({ \tiny{\tt Mura-Capua}})&\checkmark&\checkmark&\checkmark&\checkmark&\checkmark&\checkmark\\
$a_{2}$({ \tiny{\tt Cast-Capua}})&\checkmark&\checkmark&\checkmark&\checkmark&\checkmark&\checkmark\\
$a_{3}$({ \tiny{\tt Comp-Maria}})&\checkmark&\checkmark&\checkmark&\checkmark&\checkmark&\checkmark\\
$a_{4}$({ \tiny{\tt Comp-Loren}})&\checkmark&$\times$&\checkmark&$\times$&$\times$&\checkmark\\
$a_{5}$({ \tiny{\tt Comp-Grego}})&\checkmark&\checkmark&\checkmark&\checkmark&\checkmark&\checkmark\\
$a_{6}$({ \tiny{\tt Area-Duomo}})&$\times$&\checkmark&$\times$&\checkmark&\checkmark&$\times$\\
$a_{7}$({ \tiny{\tt Area-Loren}})&$\times$&\checkmark&$\times$&\checkmark&\checkmark&$\times$\\
$a_{8}$({ \tiny{\tt Teat-Neapo}})&\checkmark&\checkmark&\checkmark&\checkmark&\checkmark&\checkmark\\
$a_{9}$({ \tiny{\tt Chie-Cosma}})&$\times$&\checkmark&$\times$&\checkmark&\checkmark&$\times$\\
$a_{10}$({ \tiny{\tt Cast-D'Ovo}})&\checkmark&\checkmark&\checkmark&\checkmark&\checkmark&\checkmark\\
$a_{11}$({ \tiny{\tt Comp-Gerol}})&\checkmark&\checkmark&\checkmark&\checkmark&\checkmark&\checkmark\\
$a_{12}$({ \tiny{\tt SanG-Corvo}})&\checkmark&\checkmark&\checkmark&\checkmark&\checkmark&\checkmark\\
$a_{13}$({ \tiny{\tt SanA-Capon}})&\checkmark&\checkmark&\checkmark&\checkmark&\checkmark&\checkmark\\
$a_{14}$({ \tiny{\tt Comp-Monac}})&$\times$&$\times$&$\times$&$\times$&$\times$&$\times$\\
$a_{15}$({ \tiny{\tt Merc-Palaz}})&\checkmark&\checkmark&\checkmark&\checkmark&\checkmark&\checkmark\\
$a_{16}$({ \tiny{\tt Chie-Monac}})&\checkmark&\checkmark&\checkmark&\checkmark&\checkmark&\checkmark\\
$a_{17}$({ \tiny{\tt Comp-Mfede}})&\checkmark&\checkmark&\checkmark&\checkmark&\checkmark&\checkmark\\
$a_{18}$({ \tiny{\tt Carm-Mercato}})&$\times$&$\times$&$\times$&$\times$&$\times$&$\times$\\
$a_{19}$({ \tiny{\tt Comp-Paolo}})&\checkmark&\checkmark&\checkmark&\checkmark&\checkmark&\checkmark\\
$a_{20}$({ \tiny{\tt Vill-EbeRa}})&\checkmark&\checkmark&\checkmark&\checkmark&\checkmark&\checkmark\\
  \\ \hline\hline
 \end{tabular}
 \caption{Projects selected for different weights, budget $B(2)$ }\label{tab:selbudget2}
\end{table}

\begin{table}[htb!]
\centering
 \begin{tabular}{ccccccc} \hline\hline
  Configuration   & $F_{w_{j}^{1}}^{B(4)}$ & $F_{w_{j}^{2}}^{B(4)}$ & $F_{w_{j}^{3}}^{B(4)}$ & $F_{w_{j}^{4}}^{B(4)}$ & $F_{w_{j}^{5}}^{B(4)}$ & $F_{w_{j}^{6}}^{B(4)}$  \\ \hline
 $a_{1}$({ \tiny{\tt Mura-Capua}})&\checkmark&\checkmark&\checkmark&\checkmark&\checkmark&\checkmark\\
$a_{2}$({ \tiny{\tt Cast-Capua}})&$\times$&$\times$&$\times$&$\times$&$\times$&$\times$\\
$a_{3}$({ \tiny{\tt Comp-Maria}})&\checkmark&\checkmark&\checkmark&\checkmark&\checkmark&\checkmark\\
$a_{4}$({ \tiny{\tt Comp-Loren}})&$\times$&$\times$&$\times$&$\times$&$\times$&\checkmark\\
$a_{5}$({ \tiny{\tt Comp-Grego}})&\checkmark&\checkmark&\checkmark&\checkmark&\checkmark&\checkmark\\
$a_{6}$({ \tiny{\tt Area-Duomo}})&$\times$&$\times$&$\times$&$\times$&$\times$&$\times$\\
$a_{7}$({ \tiny{\tt Area-Loren}})&$\times$&$\times$&$\times$&$\times$&$\times$&\checkmark\\
$a_{8}$({ \tiny{\tt Teat-Neapo}})&\checkmark&\checkmark&\checkmark&\checkmark&\checkmark&\checkmark\\
$a_{9}$({ \tiny{\tt Chie-Cosma}})&$\times$&$\times$&$\times$&$\times$&$\times$&$\times$\\
$a_{10}$({ \tiny{\tt Cast-D'Ovo}})&\checkmark&\checkmark&\checkmark&\checkmark&\checkmark&\checkmark\\
$a_{11}$({ \tiny{\tt Comp-Gerol}})&\checkmark&\checkmark&\checkmark&\checkmark&\checkmark&\checkmark\\
$a_{12}$({ \tiny{\tt SanG-Corvo}})&\checkmark&\checkmark&\checkmark&\checkmark&\checkmark&$\times$\\
$a_{13}$({ \tiny{\tt SanA-Capon}})&$\times$&$\times$&$\times$&$\times$&$\times$&\checkmark\\
$a_{14}$({ \tiny{\tt Comp-Monac}})&$\times$&$\times$&$\times$&$\times$&$\times$&$\times$\\
$a_{15}$({ \tiny{\tt Merc-Palaz}})&\checkmark&\checkmark&\checkmark&\checkmark&\checkmark&$\times$\\
$a_{16}$({ \tiny{\tt Chie-Monac}})&\checkmark&\checkmark&\checkmark&\checkmark&\checkmark&\checkmark\\
$a_{17}$({ \tiny{\tt Comp-Mfede}})&\checkmark&\checkmark&\checkmark&\checkmark&\checkmark&\checkmark\\
$a_{18}$({ \tiny{\tt Carm-Mercato}})&$\times$&$\times$&$\times$&$\times$&$\times$&$\times$\\
$a_{19}$({ \tiny{\tt Comp-Paolo}})&\checkmark&\checkmark&\checkmark&\checkmark&\checkmark&$\times$\\
$a_{20}$({ \tiny{\tt Vill-EbeRa}})&$\times$&$\times$&$\times$&$\times$&$\times$&$\times$\\
  \\ \hline\hline
 \end{tabular}
 \caption{Projects selected for different weights, budget $B(4)$ }\label{tab:selbudget4}
\end{table}

\begin{table}[htb!]
\centering
 \begin{tabular}{ccccccc} \hline\hline
  Configuration   & $F_{w_{j}^{1}}^{B(6)}$ & $F_{w_{j}^{2}}^{B(6)}$ & $F_{w_{j}^{3}}^{B(6)}$ & $F_{w_{j}^{4}}^{B(6)}$ & $F_{w_{j}^{5}}^{B(6)}$ & $F_{w_{j}^{6}}^{B(6)}$  \\ \hline
$a_{1}$({ \tiny{\tt Mura-Capua}})&\checkmark&\checkmark&\checkmark&\checkmark&\checkmark&\checkmark\\
$a_{2}$({ \tiny{\tt Cast-Capua}})&$\times$&$\times$&$\times$&$\times$&$\times$&$\times$\\
$a_{3}$({ \tiny{\tt Comp-Maria}})&$\times$&$\times$&$\times$&\checkmark&\checkmark&$\times$\\
$a_{4}$({ \tiny{\tt Comp-Loren}})&$\times$&$\times$&$\times$&$\times$&$\times$&$\times$\\
$a_{5}$({ \tiny{\tt Comp-Grego}})&\checkmark&\checkmark&\checkmark&\checkmark&\checkmark&\checkmark\\
$a_{6}$({ \tiny{\tt Area-Duomo}})&$\times$&$\times$&$\times$&$\times$&$\times$&$\times$\\
$a_{7}$({ \tiny{\tt Area-Loren}})&$\times$&$\times$&$\times$&$\times$&$\times$&$\times$\\
$a_{8}$({ \tiny{\tt Teat-Neapo}})&\checkmark&\checkmark&\checkmark&$\times$&$\times$&\checkmark\\
$a_{9}$({ \tiny{\tt Chie-Cosma}})&$\times$&$\times$&$\times$&$\times$&$\times$&$\times$\\
$a_{10}$({ \tiny{\tt Cast-D'Ovo}})&\checkmark&\checkmark&\checkmark&\checkmark&\checkmark&\checkmark\\
$a_{11}$({ \tiny{\tt Comp-Gerol}})&\checkmark&\checkmark&\checkmark&\checkmark&\checkmark&\checkmark\\
$a_{12}$({ \tiny{\tt SanG-Corvo}})&\checkmark&\checkmark&\checkmark&$\times$&$\times$&\checkmark\\
$a_{13}$({ \tiny{\tt SanA-Capon}})&\checkmark&\checkmark&\checkmark&$\times$&$\times$&\checkmark\\
$a_{14}$({ \tiny{\tt Comp-Monac}})&$\times$&$\times$&$\times$&$\times$&$\times$&$\times$\\
$a_{15}$({ \tiny{\tt Merc-Palaz}})&$\times$&$\times$&$\times$&\checkmark&\checkmark&$\times$\\
$a_{16}$({ \tiny{\tt Chie-Monac}})&$\times$&$\times$&$\times$&$\times$&$\times$&$\times$\\
$a_{17}$({ \tiny{\tt Comp-Mfede}})&$\times$&$\times$&$\times$&$\times$&$\times$&$\times$\\
$a_{18}$({ \tiny{\tt Carm-Mercato}})&$\times$&$\times$&$\times$&$\times$&$\times$&$\times$\\
$a_{19}$({ \tiny{\tt Comp-Paolo}})&$\times$&$\times$&$\times$&$\times$&$\times$&$\times$\\
$a_{20}$({ \tiny{\tt Vill-EbeRa}})&$\times$&$\times$&$\times$&$\times$&$\times$&$\times$\\
  \\ \hline\hline
 \end{tabular}
 \caption{Projects selected for different weights, budget $B(6)$ }\label{tab:selbudget6}
\end{table}

 To validate the priority assigned by the {\sc{Electre Tri-nC}} method the analyst decide to conduct the selection process also for the priorities defined by the other set of weights generated by the interaction with the other specialist (see Table \ref{tab:weightstable}). Therefore, adopting the same constraints used in the previous set of simulations, the binary programming model was solved again. The results have been provided in Tables \ref{tab:selbudget2}, \ref{tab:selbudget4} and \ref{tab:selbudget6} for budgets $B(2)$, $B(4)$ and $B(6)$, respectively.

The following insights have been discussed:
\begin{itemize}[label={--}]
\item The configurations related to budget $B(1)$ generated always the same portfolio independently from the priorities generated, i.e., carrying out all the projects in our analysis apart from project $a_{14}$.
\item The configuration with budget $B(7)$ was excluded because the DM felt that with that amount of budget the choice was related only to very few projects and it was considered a scenario not worth of investigation.
\item Analyzing the portfolios generated with budgets $B(2)$ we can notice that weights $w_{j}^{1}$, $w_{j}^{3}$ and $w_{j}^{6}$ all generated the same portfolio while weights $w_{j}^{2}$, $w_{j}^{4}$ $w_{j}^{5}$ generated the same  portfolio but different from the other one. Anyway those two portfolios only differs for the presence (or the absence) of projects $a_4$, $a_6$, $a_7$, $a_9$.
\item The portfolios generated with budgets $B(4)$ are consistent for all the set of weights, apart from the set of weights $w_{j}^{6}$, i.e., {\tt{EXP-COMM}}. In particular this portfolio includes projects $a_4$, $a_7$, $a_{13}$ excluding projects $a_{12}$ and $a_{19}$. The analysts felt that those two portfolios should be again presented to the DM. The DM has commented that portfolio for weights $w_{j}^{6}$ does not include projects $a_{19}$ that in his opinion represents a crucial projects. Anyway, the second portfolio is worth of some considerations given that includes more projects.
\item Two portfolios were generated for budget $B(6)$  the first for weights $w_{j}^{1}$, $w_{j}^{2}$, $w_{j}^{3}$ and $w_{j}^{6}$ and the second one for $w_{j}^{4}$ and $w_{j}^{5}$. These two portfolios have in common all the projects apart the inclusion in the second portfolio of projects $a_{3}$ and $a_{15}$ and the exclusion of projects $a_{12}$ and $a_{13}$. Again the DM was asked to comment on those portfolios. He said that he would have preferred the first portfolio for the inclusion of project $a_8$, that will revamp an important area of the historic city center with a  boosting effect on the entire city center.

\end{itemize}

\section{Discussion (insights from practice) and major conclusions}\label{sec:insights}
\noindent

\noindent In this paper we have presented a new methodology for the prioritization and the consequent selection of a portfolio of projects in the urban context. We have showed how our methodology can be applied in a case study concerning  cultural heritage projects in the prominent cultural city of Naples. The methodology could be also used in complex contexts such as in our case study. Our methodology has been proved to be well accepted from the actors involved in the process, allowing the standpoints of different actors to be integrated in the different steps of the procedure.

The interaction with different specialists and a municipality representative  has enriched our study of multiple standpoints and the decision  has been built together with them. The interaction with them is an essential point especially when decision about public facilities have to be taken and justified to the public.

The application has been developed in the context of the cultural heritage, that needs huge investments to preserve artefacts that represent the history of the entire population. A strategy to investigate the benefits of the projects to carry out and to support the decision can be a fundamental tool for governments and associations that need to cooperate with the local business and companies in order to assure that all the standpoints will be included in the final decision.

Even if applied to a specific case study, the method can be easily extended to other fields of applications, especially considering the importance of rationalizing the expenses ever increasing in the last years.

 In the  case study considered in this article, the analysis of the options presented to the DM allowed us to formulate several insights:
\begin{itemize}[label={--}]
\item The {\sc{Electre Tri-nC}} leads to assign projects in predefined classes, allowing the DM to reduce the cognitive burden necessary to choose which projects have to be prioritized.
\item The selection process has benefited by the prioritization obtained with  {\sc{Electre Tri-nC}}. The optimization model allows us to quickly show to the DM the options available, even for different parameters, for example changing the budget. Moreover, reflections on the constraint imposed can lead to a reformulation of the model when the budget available decrease. The formulation of such constraints is very flexible and can be easily changed according to the learning process that the DM develops during the implementation of the methodology.

\item The results of the methodology have been discussed with DM and it permits to integrate his standpoint in different phases involving him in a decision that he can really feel as his own decision. The whole process was generally  well accepted.
\item The application to this  case study allowed us to test the {\sc{Electre Tri-nC}} method in a new and challenging context. The method has been   well accepted to prioritize projects to be included in a portfolio problem. This validates its flexibility in being used in such new strand of research.

\item The methodology integrates an {\sc{Electre}} method with a binary programming model. It allows to define a new interactive methodology to select feasible projects, but that can be easily extended to any  portfolio decision problems \citep{SaloEtAl11}.
\end{itemize}


Some lines for future research should include:

\begin{itemize}[label={--}]
\item Given the success of focus group in providing information, the interaction with the different participants and the decision maker should be better discussed and investigated.
\item Synergy of projects should be considered in a more systematic way.
    \item Interaction effects between criteria \citep[see][]{FigueiraEtAl09} should be incorporated in the model.
 \end{itemize}

\vspace{0.5cm}

\vfill\newpage

~~

\vspace{-3.00cm}

\renewcommand{\thesection}{\Alph{section}}
\setcounter{section}{0}
\section{Appendix}\label{Appendix:Electre-nC}
\noindent This Appendix is devoted to the presentation of {\sc{Electre Tri-nC}} method \citep{AlmeidaDiasEtAl12}, which is a generalization of the {\sc{Electre Tri-nC}} method  proposed by the same authors \citep{AlmeidaDiasEtAl10}.

\subsection{Basic data}\label{sec:basic-data}
\noindent Let $A=\{a_1,\ldots,a_i,\ldots,a_m\}$ be the set of actions. Each action is characterized according to a set of criteria, $G=\{g_1,\ldots,g_j,\ldots,g_n\}$. The performance of action $a_i$ on criterion $g_j$ is denoted by $g_j(a_i)$. Let $C=\{C_1,\ldots,C_h,\ldots,C_q\}$ the set of ordered categories, where $C_1$ is the worst and $C_q$ the best one. Each category, $C_h$ is characterized by a set of reference actions, $B_h = \{b_{h1},\ldots,b_{h\ell},\ldots,b_{h\vert B_h\vert}\}$, for $h=1,\ldots,q$. Sets $A$, $G$, and $C$ constitute our basic data.

\subsection{Modeling the imperfect knowledge: The pseudo-criterion model}\label{sec:imperfect-knowledge}
\noindent A criterion, $g_j$ is a real-valued function used for comparing two actions, $a$ and $b$, on the basis of their performances, $g_j(a)$ and $g_j(b)$, respectively \citep[see][]{Roy96, RoyBo93}. Assume that $g_j$ is to be maximized and $g_j(a) \geqslant g_j(b)$. For taking into account the imperfect knowledge of data (due to arbitrariness, uncertainty, imprecision, and ill-determination), a more sophisticated model is needed \citep[see][]{RoyEtAl14}. It is called pseudo-criterion model and can be defined as follows \citep[see][]{RoyVi84}.

\begin{definition}[Pseudo-criterion]\label{def:psedo-criterion}
A pseudo-criterion, $g_j$, is a criterion with two threshold functions associated with it: An indifference threshold function, $q_j(\cdot)$, and a preference threshold function, $p_j(\cdot)$, such that $p_j(\cdot) \geqslant q_j(\cdot) \geqslant  0$. The argument of these threshold functions may be the performance of the worst action, in which case we call them direct thresholds, or the performance of the worst action, in which case we call them inverse thresholds. For the sake of simplicity, we consider here only the case of constant thresholds, and use the notation $q_j$ and $p_j$.
\end{definition}

From this model we can derive three fundamental binary relations for criterion $g_j$ as stated in the following definition \citep[see also][]{DoumposFi18}.

\begin{definition}[\textit{per}-criterion binary preference relations]\label{def:per-binary-relations}
The following three binary relations can be derived form the model of Definition \ref{def:psedo-criterion}.
	\begin{enumerate}
    	\item per-criterion indifference relation: The  		actions $a$ and $b$ are considered indifferent  			on criterion $g_j$ ($aI_jb$), if $\vert g_j(a) - g_j(b)\vert \leqslant q_j$. This corresponds
         to the case where no significant advantage of
         one of the two actions over the other exists.
         Let $C(aIb)$ denote the coalition of criteria
         for which $aI_jb$.
        \item per-criterion strict preference relation:
        Actions $a$ is strictly preferred to $b$
        on criterion $g_j$ ($aP_jb$), if $g_j(a)
         - g_j(b) > p_j$. This corresponds
         to the case where a significant advantage of
         $a$ over $b$ exists.
         Let $C(aPb)$ denote the coalition of criteria
         for which $aP_jb$.
        \item per-criterion weak preference relation:
        Actions $a$ is weakly preferred to $b$
        on criterion $g_j$ ($aQ_jb$), if $q_j < g_j(a)
         - g_j(b) \leqslant p_j$. This corresponds
         to the case where there is not enough
         information to say if there is indifference
         between $a$ and $b$ or strict preference
         of $a$ over $b$ exists.
         Let $C(aQb)$ denote the coalition of criteria
         for which $aP_jb$.
    \end{enumerate}
\end{definition}

Let us remark that there is not an idea of intensities of preferences behind the strict and weak preference relations. Indeed, we are not modeling DM preferences, bur rather the imperfect knowledge of data.

The per-criterion binary relations of Definition \ref{def:per-binary-relations} allow us the introduce the definition of the \textit{per}-criterion outranking relation.

\begin{definition}[\textit{per}-criterion binary outranking relation]\label{def:per-outranking}
The assertion  ``$a$ outranks $b$'' ($aS_jb$) or, in other words, ``$a$ is at least as good as $b$'' is the meaning of the outranking relation $aS_jb$, which can be defined in two senses:
	\begin{enumerate}
    	\item \textit{stricto sensu}: we say that
        $aS_jb$ whenever $aI_jb$, $aQ_jb$, and
        $aP_jb$.
        \item \textit{lato sensu}  we say that
        $aS_jb$ whenever $aI_jb$, $aQ_jb$, $aP_jb$,
        but (also) when $bQ_ja$.
      \end{enumerate}
\end{definition}

{\sc{Electre}} methods are part of the family of outranking based methods and make use of the concepts in the previous definitions to construct one or several comprehensive outranking relations \citep[see][]{FigueiraEtAl16}. In the next paragraph we will present the steps leading to the construction of a single comprehensive binary relation for every ordered pair of actions $(a,b) \in A \times A$.

\subsection{Constructing an outranking relation}\label{sec:constructing-outranking}
\noindent For the construction of a crispy comprehensive outranking relation five steps should be followed.

\begin{enumerate}
    \item \textit{The power of the concordant
    coalition.} The power of the concordant
    coalition of the assertion ``$a$ outranks
    $b$ can be interpreted as the strength
    of the arguments favoring action $a$ over $b$.
    It takes into account the weights of criteria
    belonging to the concordant coalition
    \textit{stricto sensu}, i.e., $w_j$ such
    that $g_j \in C(a\{I,Q,P\}b)$, in its
    totality, plus a proportion, $phi_j$, of the
    weights for the criteria belonging to the
    ambiguity zone, i.e., $\phi_j w_j$ for
    all $g_j \in C(bQa)$. The power of the
    concordant coalition is this modeled through
    a comprehensive concordant index as follows:

    \begin{equation}\label{eq:concordant-index}
    	{\displaystyle c(a,b) = \sum_{\{j\,:\,g_j
        \in C(a\{I,Q,P)b\}}w_j +\sum_{\{j\,:\,g_j
       \in C(bQa)\}}\phi_jw_j},
    \end{equation}
    \noindent where
    \[
    	{\displaystyle \phi_j = \frac{g_j(a)
        - g_j(b) + p_j}{p_j - q_k} \; \in \;
        [0,1].}
    \]
    In presence of interaction effects between criteria, this index can be extended as in \cite{FigueiraEtAl09}.
	\item \textit{The effect of the discordant
    criteria.} The effect of a discordant
    criteria $g_j$ reflects the  opposing power of
    such criterion, and it makes use of another
    preference parameter called veto threshold,
    which may vary (directly or inversely) or
    be constant as the indifference and preference
    thresholds.  Let us consider such a veto
    threshold and constant and denote it by $v_j$.
    The opposing power of this criterion can be
    modeled through a per-criterion discordance as
    follows.

    \begin{equation}\label{eq:discordant-index}
    d_j(a,b) = \left\{
    	\begin{array}{ccl}
         1 & \mbox{if} & g_j(a) - g_j(b) < - v_j,\\
         \frac{g_j(a) - g_j(b) + p_j}{p_j - v_j},
         & \mbox{if} &
        -v_j \leqslant g_j(a) - g_j(b) < - p_j, \\
         0 & \mbox{if} &
         g_j(a) - g_j(b) \geqslant - p_j.\\
       \end{array}
        \right.
    \end{equation}
	\item \textit{The credibility degree of an
    outranking relation.} The degree in which $a$
    outranks $b$ can be modeled through the
    following formula.
    \begin{equation}\label{eq:credibility-index}
    	{\displaystyle \sigma(a,b) =
        \prod_{j=1}^{n}}T_j(a,b),
    \end{equation}
    \noindent where
    \[
    T_j(a,b) = \left\{
    	\begin{array}{cl}
         \frac{1 - d_j(a,b)}{1 - c(a,b)}
         & \mbox{if}  \;\, d_j(a,b) > c(a,b), \\
         0 & \mbox{otherwise.}
       \end{array}
        \right.
    \]

	\item \textit{The category credibility indices.}  These indices model the credibility degree of an action with respect to a set and \textit{vice-versa}. The justification for the choice of the {\tt max} operator can be seen in \cite{AlmeidaDiasEtAl12}.
    \begin{equation}
    \begin{array}{l}
     	{\displaystyle \sigma(a,B_h) =
        \max_{\ell=1,\ldots,
        \vert B_h \vert}\Big\{\sigma(a,b_{h
        \ell})\Big\}} \\
        {\displaystyle \sigma(B_h,a) =
        \max_{\ell=1,\ldots,
        \vert B_h \vert}\Big\{\sigma(b_{h
        \ell},a)\Big\}} \\
    \end{array}
    \end{equation}
    At the end of this process we defined a fuzzy relation for each ordered pair $(a,B_k)$ and $(B_h,a)$, for $h=1,\ldots,q$.
	\item \textit{The comprehensive outranking
    relations.} A comprehensive crisp outranking relation is obtained after considering a $\lambda-$cutting off level. We say that ``$a$ comprehensively outranks $B_h$'', denoted by $aSB_h$, if $\sigma(a,B_h) \geqslant \lambda$. Otherwise, ``$a$ does not outrank $B_h$'', denoted by $\mbox{not}(aSB_h)$. The outranking relation should also be checked in the reverse situation, i.e., for the ordered pair $(B_h,a)$. For situations may occur:
    \begin{enumerate}
    	\item $(aSB_h)$ and $(B_hSa)$: indifference
        between $a$ and $b$ ($aIb$);
        \item  $(aSB_h)$ and $\mbox{not}(B_hSa)$:
        preference of $a$ over $b$ ($a \succ b$);
        \item  $(B_hSa)$ and $\mbox{not}(aSB_h)$:
        preference of $b$ over $a$ ($b \succ a$);
        \item $\mbox{not}(aSB_h)$ and
        $\mbox{not}(B_hSa)$: incomparability
        between $a$ and $b$ ($aRb$).
    \end{enumerate}
\end{enumerate}
The building of comprehensive binary relations finishes at this point. Next subsection introduces the procedures allowing to exploit this information.

\subsection{Exploiting the  relation: The assignment procedures}\label{sec:constructing-outranking}
\noindent Before introducing the two assignment procedures for the method, it is necessary to present the selecting function.

\begin{equation}\label{def:selecting_function}
	{\displaystyle \rho(a,B_h) = \min\Big\{ \sigma(a,B_h), \; \sigma(B_h,a) \Big\}}.
\end{equation}

Its justification can be found in \cite{AlmeidaDiasEtAl12}. The two procedures of {\sc{Electre Tri-nC}}, which exploit the outranking relations and provides the assignments to each action can be presented as follows. Consider two dummy profiles, $B_0$ ($B_{q+1}$), which is dominated or dominates  all the potential actions to be assigned.

\begin{definition}[Descending procedure]\label{def:descending}
Consider $\lambda \in [0.5,\,1]$. Decrease $h$ from $(q + 1)$ until the first value, $t$, such that $\sigma(a, B_t) \geqslant \lambda$. Then,
	\begin{enumerate}
	\item For $t = q$, consider $C_q$ as a possible
    category to assign action $a$.
    \item For $0 < t < q$, if $\rho(a,B_t) > \rho(a, B_{t+1})$, then consider $C_t$ as a possible category to assign $a$; otherwise, consider $C_{t+1}$.
    \item For $t = 0$, consider $C_1$ as a possible category to assign $a$.
	\end{enumerate}
\end{definition}

\begin{definition}[Ascending procedure]\label{def:descending}
Consider $\lambda \in [0.5,\,1]$. Increase $h$ from $0$ until the first value, $k$, such that $\sigma(B_k,a) \geqslant \lambda$. Then,
	\begin{enumerate}
	\item For $k = 1$, consider $C_1$ as a possible
    category to assign action $a$.
    \item For $1 < k < q + 1$, if $\rho(a, B_k) >
    \rho(a, B_{k,1})$, then consider $C_k$ as a
    possible category to assign $a$; otherwise,
    select $C_{k,1}$.
    \item For $k = q + 1$, consider $C_q$ as a
    possible category to assign $a$.
	\end{enumerate}
\end{definition}

\vfill\newpage

~~

\vspace{-3.00cm}

\renewcommand{\thesection}{\Alph{section}}
\setcounter{section}{1}
\section{Appendix}\label{sec:Appendix_Data}
\noindent

\vspace{1.25cm}

{\scriptsize
\begin{table}[h!]
\centering
 \begin{tabular}{rccccccccccc} \hline\hline
  \small{Criteria}  & $g_1$ & $g_2$ & $g_3$ & $g_4$ &
  $g_5$ & $g_6$ & $g_7$ & $g_8$  & Cost  \\
  \small{Label}    & {\scriptsize {\tt CON-COMP}} & {\scriptsize {\tt CON-USAB}} & {\scriptsize {\tt PRO-CRAF}} & {\scriptsize {\tt PRO-BUSI}} & {\scriptsize {\tt PRO-TOUR}} & {\scriptsize {\tt ENV-MAIN}} &  {\scriptsize {\tt SOC-CULT}} & {\scriptsize {\tt SOC-COHE}}  & \\ \hdashline
  \small{Unit}     & Qual.  & \% & Qual.  & Qual.
  & Qual.  & $m^2$  & Qual.  & Qual.  & (K\euro) \\
  \small{Direction} & $\max$ & $\max$ & $\max$ & $\max$
  & $\max$ & $\max$ & $\max$ & $\max$  & \\ \hdashline
  \small{min perf.}& L(1)  & 25  & L-M(2)  & L(1)  & L(1)  & 0  & L-L(1) & L(1)  & \\
  \small{max perf.} & VH(4) & 100 & VH-L(13) & VH(4) & VH(4) & 21000 &  VH-H(15) & VH(4)  & \\ \hline\hline
$a_{1}$({ \tiny{\tt Mura-Capua}})
&
VH(4)&80&H-L(9)&M(2)&H(3)&5650&H-M(10)&H(3) & 1500 \\
$a_{2}$({ \tiny{\tt Cast-Capua}})&VH(4)&25&L-H(3)&M(2)&H(3)&3500&H-H(11)&VH(4) & 5000 \\
$a_{3}$({ \tiny{\tt Comp-Maria}})&L(1)&80&H-H(11)&H(3)&H(3)&9750&VH-VH(16)&H(3) & 7000 \\
$a_{4}$({ \tiny{\tt Comp-Loren}})&H(3)&20&L-VH(4)&L(1)&M(2)&2000&H-M(10)&VH(4) & 3000 \\
$a_{5}$({ \tiny{\tt Comp-Grego}})&H(3)&30&M-VH(8)&L(1)&H(3)&2400&H-M(10)&VH(4) & 1100 \\
$a_{6}$({ \tiny{\tt Area-Duomo}})&H(3)&50&L-L(1)&L(1)&H(3)&0&M-M(6)&L(1) & 1500 \\
$a_{7}$({ \tiny{\tt Area-Loren}})&H(3)&25&L-L(1)&L(1)&H(3)&0&H-H(11)&M(2) & 1000 \\
$a_{8}$({ \tiny{\tt Teat-Neapo}})&H(3)&60&M-H(7)&M(2)&VH(4)&3600&VH-VH(16)&H(3) &  6000 \\
$a_{9}$({ \tiny{\tt Chie-Cosma}})&H(3)&100&L-M(2)&L(1)&L(1)&1700&H-H(11)&VH(4) & 900 \\
$a_{10}$({ \tiny{\tt Cast-D'Ovo}})&H(3)&60&M-M(6)&M(2)&VH(4)&0&H-H(11)&M(2) & 1500 \\
$a_{11}$({ \tiny{\tt Comp-Gerol}})&VH(4)&100&H-H(11)&M(2)&H(3)&12000&VH-VH(16)&H(3) & 7700 \\
$a_{12}$({ \tiny{\tt SanG-Corvo}})&H(3)&70&L-M(2)&L(1)&M(2)&1700&H-M(10)&M(2) & 400 \\
$a_{13}$({ \tiny{\tt SanA-Capon}})&H(3)&50&L-M(2)&L(1)&H(3)&0&VH-H(15)&M(2) & 1000 \\
$a_{14}$({ \tiny{\tt Comp-Monac}})&M(2)&70&M-H(7)&M(2)&H(3)&21000&H-M(10)&VH(4) & 13600 \\
$a_{15}$({ \tiny{\tt Merc-Palaz}})&H(3)&100&VH-L(13)&VH(4)&L(1)&360&L-L(1)&VH(4) & 500 \\
$a_{16}$({ \tiny{\tt Chie-Monac}})&H(3)&50&L-VH(4)&L(1)&M(2)&200&H-L(9)&M(2) & 500 \\
$a_{17}$({ \tiny{\tt Comp-Mfede}})&H(3)&90&M-H(7)&L(1)&M(2)&3500&H-L(9)&VH(4) & 2000 \\
$a_{18}$({ \tiny{\tt Carm-Merca}})&M(2)&30&H-L(9)&VH(4)&L(1)&1500&M-M(6)&H(3) & 4000 \\
$a_{19}$({ \tiny{\tt Comp-Paolo}})&H(3)&50&L-H(3)&M(2)&H(3)&5200&VH-H(15)&H(3) & 4000 \\
$a_{20}$({ \tiny{\tt Vill-EbeRa}})&M(2)&100&M-M(6)&H(3)&VH(4)&600&M-H(7)&M(2) & 4100 \\
\hline\hline
 \end{tabular}
 \caption{Performance table}\label{tab:PerformanceTabl}
\end{table}
}

\vfill\newpage

\vspace{1.5cm}

{\tiny
\begin{table}[h!]
\centering
 \begin{tabular}{ccccccccc} \hline\hline
 {\footnotesize{Criteria}}  & $g_1$ & $g_2$ & $g_3$ & $g_4$ &
  $g_5$ & $g_6$ & $g_7$ & $g_8$  \\
  Label   & {\scriptsize {\tt CON-COMP}} & {\scriptsize {\tt CON-USAB}} & {\scriptsize {\tt PRO-CRAF}} & {\scriptsize {\tt PRO-BUSI}} & {\scriptsize {\tt PRO-TOUR}} & {\scriptsize {\tt ENV-MAIN}} &  {\scriptsize {\tt SOC-CULT}} & {\scriptsize {\tt SOC-COHE}}  \\ \hline
 $q_j(b)$ & $--$ & \scriptsize {$0.10g_2(b)+13.00$} & $1$ & $--$ & $--$ & \scriptsize {$0.10g_6(b)+397.73$} & $1$ & $--$ \\
  $p_j(b)$ & $--$ & \scriptsize{$0.00g_2(b)+30.00$} & $3$ & $--$ & $--$ & \scriptsize{$0.21g_6(b)+795.46$} & $3$ & $--$ \\ \hdashline
  $v_j(b)$ & $3$ & \scriptsize {$0.000g_2(b)+110.00$}  & $5$ & $3$ & $3$ & \scriptsize {$0.00g_6(b)+60000.00$} & $5$ & $3$  \\ \hline\hline
 \end{tabular}
 \caption{Discriminating and veto thresholds}\label{tab:veto}
\end{table}
}

\vspace{1.5cm}

\begin{table}[h!]
\centering
 \begin{tabular}{ccccccc}  \hline\hline
  Functions  &  {\footnotesize{\thead{Tourist\\ Facilities}}}   &{ \footnotesize{\thead{ Archaeological and \\ Museum  sites}}}&{ \footnotesize{\thead{ Students and elderly \\ housing}}}&{ \footnotesize{\thead{ Leisure\\ Activities}}} &{ \footnotesize{\thead { Record Offices \\Archives}}}&{ \footnotesize{\thead{ Citizenship \\facilities}}}   \\
  Label   & $U_{1}$ & $U_{2}$ & $U_{3}$ & $U_{4}$ & $U_{5}$ & $U_{6}$   \\ \hline
$a_{1}$({ \tiny{\tt Mura-Capua}})&\checkmark&$\times$&$\times$&$\times$&$\times$&\checkmark\\
$a_{2}$({ \tiny{\tt Cast-Capua}})&\checkmark&\checkmark&$\times$&$\times$&$\times$&\checkmark\\
$a_{3}$({ \tiny{\tt Comp-Maria}})&\checkmark&\checkmark&\checkmark&\checkmark&$\times$&\checkmark\\
$a_{4}$({ \tiny{\tt Comp-Loren}})&$\times$&$\times$&$\times$&$\times$&\checkmark&\checkmark\\
$a_{5}$({ \tiny{\tt Comp-Grego}})&\checkmark&\checkmark&\checkmark&$\times$&$\times$&\checkmark\\
$a_{6}$({ \tiny{\tt Area-Duomo}})&$\times$&$\times$&\checkmark&$\times$&$\times$&$\times$\\
$a_{7}$({ \tiny{\tt Area-Loren}})&$\times$&$\times$&\checkmark&$\times$&$\times$&$\times$\\
$a_{8}$({ \tiny{\tt Teat-Neapo}})&$\times$&$\times$&\checkmark&\checkmark&$\times$&$\times$\\
$a_{9}$({ \tiny{\tt Chie-Cosma}})&$\times$&$\times$&$\times$&\checkmark&$\times$&\checkmark\\
$a_{10}$({ \tiny{\tt Cast-D'Ovo}})&$\times$&$\times$&$\times$&\checkmark&$\times$&\checkmark\\
$a_{11}$({ \tiny{\tt Comp-Gerol}})&\checkmark&\checkmark&\checkmark&\checkmark&\checkmark&\checkmark\\
$a_{12}$({ \tiny{\tt SanG-Corvo}})&$\times$&$\times$&$\times$&$\times$&\checkmark&$\times$\\
$a_{13}$({ \tiny{\tt SanA-Capon}})&\checkmark&\checkmark&$\times$&$\times$&\checkmark&$\times$\\
$a_{14}$({ \tiny{\tt Comp-Monac}})&$\times$&$\times$&$\times$&\checkmark&$\times$&\checkmark\\
$a_{15}$({ \tiny{\tt Merc-Palaz}})&$\times$&$\times$&$\times$&$\times$&$\times$&\checkmark\\
$a_{16}$({ \tiny{\tt Chie-Monac}})&$\times$&$\times$&$\times$&$\times$&$\times$&\checkmark\\
$a_{17}$({ \tiny{\tt Comp-Mfede}})&$\times$&$\times$&\checkmark&\checkmark&$\times$&\checkmark\\
$a_{18}$({ \tiny{\tt Carm-Mercato}})&$\times$&$\times$&$\times$&$\times$&$\times$&\checkmark\\
$a_{19}$({ \tiny{\tt Comp-Paolo}})&\checkmark&\checkmark&\checkmark&\checkmark&$\times$&$\times$\\
$a_{20}$({ \tiny{\tt Vill-EbeRa}})&\checkmark&$\times$&$\times$&\checkmark&$\times$&\checkmark\\

  \\ \hline\hline
 \end{tabular}
 \caption{Functions for each project}\label{tab:Functions}
\end{table}

\vfill\newpage

\vspace{1.5cm}

\noindent In the following Tables we report the ranking of the criteria provided by each specialist. The column ``Number of card'' indicates the number of card inserted by the specialist between the considered level and the previous one.

\begin{table}[h!]
\centering
 \begin{tabular}{ccc} \hline\hline
Rank & Criteria & Number of cards \\
1 & \{$g_2$, $g_4$, $g_6$\} & 0 \\
2 & \{$g_1$, $g_7$\} & 1 \\
3 & \{$g_3$, $g_5$, $g_8$\} & 2 \\
\hline
 \end{tabular}
 \caption{Specialist in preservation of tangible cultural heritage standpoint (Ratio = 3)}\label{tab:expconv}
\end{table}

\begin{table}[h!]
\centering
 \begin{tabular}{ccc} \hline\hline
Rank & Criteria & Number of cards \\
1 & \{$g_3$, $g_4$, $g_7$, $g_{8}$\} & 0 \\
2 & \{$g_1$, $g_2$, $g_6$\} & 2 \\
3 & \{ $g_5$\} & 5 \\
\hline
 \end{tabular}
 \caption{ Specialist in promotion of the traditional craftsmanship and local products (Ratio = 10)}\label{tab:exprom}
\end{table}

\begin{table}[h!]
\centering
 \begin{tabular}{ccc} \hline\hline
Rank & Criteria & Number of cards \\
1 & \{$g_6$\} & 0 \\
2 & \{$g_8$\} & 0 \\
3 & \{$g_2$\} & 0 \\
4 & \{$g_3$, $g_4$, $g_5$\} & 0 \\
5 & \{ $g_7$\} & 0 \\
6 & \{$g_1$\} & 0 \\
\hline
 \end{tabular}
 \caption{Specialist in  quality of urban environment standpoint (Ratio = 5)}\label{tab:expurbe}
\end{table}

\begin{table}[h!]
\centering
 \begin{tabular}{ccc} \hline\hline
Rank & Criteria & Number of cards \\
1 & \{$g_1$\} & 0 \\
2 & \{$g_6$\} & 0\\
3 & \{$g_3$, $g_8$\} & 0 \\
1 & \{$g_7$\} & 1 \\
2 & \{$g_2$\} & 2 \\
3 & \{$g_4$, $g_5$\} & 0 \\
\hline
 \end{tabular}
 \caption{Specialist in social benefits for the community standpoint (Ratio = 5)}\label{tab:expcomm}

\end{table}

\vfill\newpage

\section{Appendix}\label{sec:Appendix_Model}

\begin{table}[h!]
\centering
 \begin{tabular}{rcccccccccc} \hline\hline
  Specialist  & $g_1$ & $g_2$ & $g_3$ & $g_4$ &
  $g_5$ & $g_6$\\
 \hline\hline
$a_{1}$({ \tiny{\tt Mura-Capua}})&16&42&16&56&72&36\\
$a_{2}$({ \tiny{\tt Cast-Capua}})&16&42&16&56&72&36\\
$a_{3}$({ \tiny{\tt Comp-Maria}})&160&378&160&504&648&360\\
$a_{4}$({ \tiny{\tt Comp-Loren}})&16&6&16&4&6&36\\
$a_{5}$({ \tiny{\tt Comp-Grego}})&16&42&16&56&72&36\\
$a_{6}$({ \tiny{\tt Area-Duomo}})&1&1&1&1&1&1\\
$a_{7}$({ \tiny{\tt Area-Loren}})&2&2&2&2&1&2\\
$a_{8}$({ \tiny{\tt Teat-Neapo}})&160&378&160&504&648&360\\
$a_{9}$({ \tiny{\tt Chie-Cosma}})&2&6&2&4&6&2\\
$a_{10}$({ \tiny{\tt Cast-D'Ovo}})&16&42&16&56&72&36\\
$a_{11}$({ \tiny{\tt Comp-Gerol}})&480&1134&480&1512&1944&1080\\
$a_{12}$({ \tiny{\tt SanG-Corvo}})&2&6&2&4&6&2\\
$a_{13}$({ \tiny{\tt SanA-Capon}})&2&2&2&4&3&12\\
$a_{14}$({ \tiny{\tt Comp-Monac}})&16&42&16&56&72&36\\
$a_{15}$({ \tiny{\tt Merc-Palaz}})&2&6&2&28&36&2\\
$a_{16}$({ \tiny{\tt Chie-Monac}})&2&6&2&4&6&12\\
$a_{17}$({ \tiny{\tt Comp-Mfede}})&16&42&16&56&72&36\\
$a_{18}$({ \tiny{\tt Carm-Mercato}})&2&6&2&4&6&2\\
$a_{19}$({ \tiny{\tt Comp-Paolo}})&16&42&16&56&72&36\\
$a_{20}$({ \tiny{\tt Vill-EbeRa}})&16&42&16&56&72&36\\
\hline\hline
 \end{tabular}
 \caption{Objective Function: $c_h$ coefficients}\label{Tab:chcoeff}
\end{table}

\vspace{1.5cm}

\textbf{Constraints}:

\begin{enumerate}

\item Constraints projects on the ``decumano'', constraint (\ref{con:decumano}):
$x_{2}+x_{3}+x_{7}+x_{8}+x_{11}+x_{19}\geqslant 4$.

\item Constraints synergy on ``insula'', constraints (\ref{con:siner1}) to (\ref{con:siner4}):
$x_{47}\geqslant x_{4}+x_{7}-1;$
$x_{4} \geqslant x_{47};$
$x_{7} \geqslant x_{47};$
$x_{819}\geqslant x_8+x_{19}-1;$
$x_{8}\geqslant x_{819};$
$x_{19}\geqslant x_{819};$
$x_{47}+x_{819}\geqslant 1.$

\item Constraints function $U_3$ (Student and elderly accommodation), constraint: (\ref{con:function1})
$x_{3}+x_{11}+x_{17}+x_{19}\geqslant 3.$

\item Constraints distribution of functions $U_1$ and $U_6$, constraints (\ref{con:function2})-(\ref{con:function3}):
$x_{13}+x_{19}\geqslant 1-3y_{1};$
$x_{12}\geqslant 1-3y_{1};$
$x_{1}+x_{2}+x_{11}\geqslant 1-3y_{2};$
$x_5\geqslant 1-3y_3;$
$x_3+x_4\geqslant 1-3y_3;$
$x_3\geqslant 1-3y_4;$
$x_3+x_4\geqslant 1-3y_4;$
$y_1+y_2+y_3+y_4 \leqslant 1.$

\end{enumerate}

\begin{landscape}

\section{Appendix}\label{sec:datisecondi}

\vspace{2cm}

\begin{table}[htb!]
\centering
 \begin{tabular}{ccccccccc} \hline\hline
  Representative  & $g_1$ & $g_2$ & $g_3$ & $g_4$ &
  $g_5$ & $g_6$ & $g_7$ & $g_8$  \\
  Actions   & {\scriptsize {\tt CON-COMP}} & {\scriptsize {\tt CON-USAB}} & {\scriptsize {\tt PRO-CRAF}} & {\scriptsize {\tt PRO-BUSI}} & {\scriptsize {\tt PRO-TOUR}} & {\scriptsize {\tt ENV-MAIN}} &  {\scriptsize {\tt SOC-CULT}} & {\scriptsize {\tt SOC-COHE}}  \\ \hline
  $b_{1,1}$ & 1 & 30 & 3  & 1 & 1 & 200  & 4  & 1\\ \hdashline
  $b_{2,1}$ & 2 & 40 & 8  & 2 & 2 & 12000& 9  & 2\\
  $b_{2,2}$ & 2 & 45 & 7  & 2 & 2 & 10000& 8  & 2\\
  $b_{2,3}$ & 2 & 50 & 9  & 2 & 2 & 15000& 7  & 2\\ \hdashline
  $b_{3,1}$ & 3 & 75 & 12 & 3 & 3 & 25000& 11 & 2\\ \hdashline
  $b_{4,1}$ & 4 & 85 & 16 & 4 & 4 & 45000& 15 & 4\\
  $b_{4,2}$ & 4 & 95 & 14 & 4 & 4 & 40000& 16 & 4\\ \hline\hline
 \end{tabular}
 \caption{tab: Reference actions analyst attempt}
\end{table}

\vspace{2cm}

\begin{table}[h!]
\centering
 \begin{tabular}{ccccccccc} \hline\hline
  Criteria  & $g_1$ & $g_2$ & $g_3$ & $g_4$ &
  $g_5$ & $g_6$ & $g_7$ & $g_8$  \\
  Label   & {\scriptsize {\tt CON-COMP}} & {\scriptsize {\tt CON-USAB}} & {\scriptsize {\tt PRO-CRAF}} & {\scriptsize {\tt PRO-BUSI}} & {\scriptsize {\tt PRO-TOUR}} & {\scriptsize {\tt ENV-MAIN}} &  {\scriptsize {\tt SOC-CULT}} & {\scriptsize {\tt SOC-COHE}}  \\ \hline
  $q_j(b)$ & $--$ & $0.030g_2(b)+1.00$ & $1$ & $--$ & $--$ & $0.045g_6(b)+454.550$ & $1$ & $--$ \\
  $p_j(b)$ & $--$ & $0.070g_2(b)+2.00$  & $3$ & $--$ & $--$ & $0.091g_6(b)+909.090$ & $3$ & $--$ \\ \hdashline
  $v_j(b)$ & $3$ & $0.090g_2(b)+3.00$  & $5$ & $3$ & $3$ & $0.120g_6(b)+1200.00$ & $5$ & $3$  \\ \hline\hline
 \end{tabular}
 \caption{Discriminating and veto thresholds}
\end{table}\label{tab:veto2}

\end{landscape}

\bibliographystyle{plainnat}
\bibliography{Bib_Cultural}

\begin{thebibliography}{58}
\expandafter\ifx\csname natexlab\endcsname\relax\def\natexlab#1{#1}\fi
\expandafter\ifx\csname url\endcsname\relax
  \def\url#1{\texttt{#1}}\fi
\expandafter\ifx\csname urlprefix\endcsname\relax\def\urlprefix{URL }\fi
\providecommand{\eprint}[2][]{\url{#2}}
\providecommand{\bibinfo}[2]{#2}
\ifx\xfnm\relax \def\xfnm[#1]{\unskip,\space#1}\fi
\bibitem[{Abastante et~al.(2018)Abastante, Corrente, Greco, Ishizaka and
  Lami}]{abastante2018choice}
\bibinfo{author}{Abastante, F.}, \bibinfo{author}{Corrente, S.},
  \bibinfo{author}{Greco, S.}, \bibinfo{author}{Ishizaka, A.},
  \bibinfo{author}{Lami, I.M.}, \bibinfo{year}{2018}.
\newblock \bibinfo{title}{Choice architecture for architecture choices:
  Evaluating social housing initiatives putting together a parsimonious ahp
  methodology and the choquet integral}.
\newblock \bibinfo{journal}{Land Use Policy} \bibinfo{volume}{78},
  \bibinfo{pages}{748--762}.
\bibitem[{Agudelo-Vera et~al.(2012)Agudelo-Vera, Leduc, Mels and
  Rijnaarts}]{agudelo2012harvesting}
\bibinfo{author}{Agudelo-Vera, C.M.}, \bibinfo{author}{Leduc, W.R.},
  \bibinfo{author}{Mels, A.R.}, \bibinfo{author}{Rijnaarts, H.H.},
  \bibinfo{year}{2012}.
\newblock \bibinfo{title}{Harvesting urban resources towards more resilient
  cities}.
\newblock \bibinfo{journal}{Resources, Conservation and Recycling}
  \bibinfo{volume}{64}, \bibinfo{pages}{3--12}.
\bibitem[{Ahern(2011)}]{ahern2011fail}
\bibinfo{author}{Ahern, J.}, \bibinfo{year}{2011}.
\newblock \bibinfo{title}{From fail-safe to safe-to-fail: Sustainability and
  resilience in the new urban world}.
\newblock \bibinfo{journal}{Landscape and Urban Planning}
  \bibinfo{volume}{100}, \bibinfo{pages}{341--343}.
\bibitem[{Almeida-Dias et~al.(2010)Almeida-Dias, Figueira and
  Roy}]{AlmeidaDiasEtAl10}
\bibinfo{author}{Almeida-Dias, J.}, \bibinfo{author}{Figueira, J.R.},
  \bibinfo{author}{Roy, B.}, \bibinfo{year}{2010}.
\newblock \bibinfo{title}{{\sc{Electre Tri-C}}: {A} multiple criteria sorting
  method based on characteristic reference actions}.
\newblock \bibinfo{journal}{European Journal of Operational Research}
  \bibinfo{volume}{204}, \bibinfo{pages}{565--580}.
\bibitem[{Almeida-Dias et~al.(2012)Almeida-Dias, Figueira and
  Roy}]{AlmeidaDiasEtAl12}
\bibinfo{author}{Almeida-Dias, J.}, \bibinfo{author}{Figueira, J.R.},
  \bibinfo{author}{Roy, B.}, \bibinfo{year}{2012}.
\newblock \bibinfo{title}{A multiple criteria sorting method where each
  category is characterized by several reference actions: {T}he
  {E}{\sc{lectre}} {T}{\sc{ri}}-{C} method}.
\newblock \bibinfo{journal}{European Journal of Operational Research}
  \bibinfo{volume}{217}, \bibinfo{pages}{567--579}.
\bibitem[{Barbati et~al.(2018)Barbati, Greco, Kadzi{\'n}ski and
  S{\l}owi{\'n}ski}]{barbati2018}
\bibinfo{author}{Barbati, M.}, \bibinfo{author}{Greco, S.},
  \bibinfo{author}{Kadzi{\'n}ski, M.}, \bibinfo{author}{S{\l}owi{\'n}ski, R.},
  \bibinfo{year}{2018}.
\newblock \bibinfo{title}{Optimization of multiple satisfaction levels in
  portfolio decision analysis}.
\newblock \bibinfo{journal}{Omega, The International Journal of Management
  Science} \bibinfo{volume}{78}, \bibinfo{pages}{192--204}.
\bibitem[{Blake(2000)}]{Blake00}
\bibinfo{author}{Blake, J.}, \bibinfo{year}{2000}.
\newblock \bibinfo{title}{On defining the cultural heritage}.
\newblock \bibinfo{journal}{International \& Comparative Law Quarterly}
  \bibinfo{volume}{49}, \bibinfo{pages}{61--85}.
\bibitem[{Bottero et~al.(2015)Bottero, Ferretti, Figueira, Greco and
  Roy}]{BotteroEtAl15}
\bibinfo{author}{Bottero, M.}, \bibinfo{author}{Ferretti, V.},
  \bibinfo{author}{Figueira, J.R.}, \bibinfo{author}{Greco, S.},
  \bibinfo{author}{Roy, B.}, \bibinfo{year}{2015}.
\newblock \bibinfo{title}{Dealing with a multiple criteria environmental
  problem with interaction effects between criteria through an extension of the
  {\sc{electre iii}} method}.
\newblock \bibinfo{journal}{European Journal of Operational Research}
  \bibinfo{volume}{245}, \bibinfo{pages}{837--850}.
\bibitem[{Bottero et~al.(2018)Bottero, Ferretti, Figueira, Greco and
  Roy}]{BotteroEtAl18}
\bibinfo{author}{Bottero, M.}, \bibinfo{author}{Ferretti, V.},
  \bibinfo{author}{Figueira, J.R.}, \bibinfo{author}{Greco, S.},
  \bibinfo{author}{Roy, B.}, \bibinfo{year}{2018}.
\newblock \bibinfo{title}{On the {C}hoquet multiple criteria preference
  aggregation model: {T}heoretical and practical insights from a real-world
  application}.
\newblock \bibinfo{journal}{European Journal of Operational Research}
  \bibinfo{volume}{245}, \bibinfo{pages}{120--140}.
\bibitem[{B{\"u}y{\"u}k{\"o}zkan et~al.(2018)B{\"u}y{\"u}k{\"o}zkan,
  Feyzio{\u{g}}lu and G{\"o}{\c{c}}er}]{buyukozkan2018selection}
\bibinfo{author}{B{\"u}y{\"u}k{\"o}zkan, G.}, \bibinfo{author}{Feyzio{\u{g}}lu,
  O.}, \bibinfo{author}{G{\"o}{\c{c}}er, F.}, \bibinfo{year}{2018}.
\newblock \bibinfo{title}{Selection of sustainable urban transportation
  alternatives using an integrated intuitionistic fuzzy choquet integral
  approach}.
\newblock \bibinfo{journal}{Transportation Research Part D: Transport and
  Environment} \bibinfo{volume}{58}, \bibinfo{pages}{186--207}.
\bibitem[{{\c{C}}a{\u{g}}lar and G{\"u}rel(2017)}]{CauglarGu17}
\bibinfo{author}{{\c{C}}a{\u{g}}lar, M.}, \bibinfo{author}{G{\"u}rel, S.},
  \bibinfo{year}{2017}.
\newblock \bibinfo{title}{Public r\&d project portfolio selection problem with
  cancellations}.
\newblock \bibinfo{journal}{OR Spectrum} \bibinfo{volume}{39},
  \bibinfo{pages}{659--687}.
\bibitem[{Cerreta and Panaro(2017)}]{cerreta2017perceived}
\bibinfo{author}{Cerreta, M.}, \bibinfo{author}{Panaro, S.},
  \bibinfo{year}{2017}.
\newblock \bibinfo{title}{From perceived values to shared values: A
  {M}ulti-{S}takeholder {S}patial {D}ecision {A}nalysis ({M}-{S}{S}{D}{A}) for
  resilient landscapes}.
\newblock \bibinfo{journal}{Sustainability} \bibinfo{volume}{9},
  \bibinfo{pages}{1113}.
\bibitem[{Cerreta and Toro(2010)}]{cerreta2010}
\bibinfo{author}{Cerreta, M.}, \bibinfo{author}{Toro, P.D.},
  \bibinfo{year}{2010}.
\newblock \bibinfo{title}{Integrated spatial assessment for a creative
  decision-making process: A combined methodological approach to strategic
  environmental assessment}.
\newblock \bibinfo{journal}{International Journal of Sustainable Development}
  \bibinfo{volume}{13}, \bibinfo{pages}{17--30}.
\bibitem[{Cranmer et~al.(2018)Cranmer, Baker, Liesi{\"o} and
  Salo}]{CranmerEtAl18}
\bibinfo{author}{Cranmer, A.}, \bibinfo{author}{Baker, E.},
  \bibinfo{author}{Liesi{\"o}, J.}, \bibinfo{author}{Salo, A.},
  \bibinfo{year}{2018}.
\newblock \bibinfo{title}{A portfolio model for siting offshore wind farms with
  economic and environmental objectives}.
\newblock \bibinfo{journal}{European Journal of Operational Research}
  \bibinfo{volume}{267}, \bibinfo{pages}{304 -- 314}.
\bibitem[{Doumpos and Figueira(2018)}]{DoumposFi18}
\bibinfo{author}{Doumpos, M.}, \bibinfo{author}{Figueira, J.R.},
  \bibinfo{year}{2018}.
\newblock \bibinfo{title}{A multicriteria outranking approach for modeling
  corporate credit ratings: {A}n application of the {\sc{electre tri-nc}}
  method}.
\newblock \bibinfo{journal}{{\sc{Omega}}, The International Journal of
  Management Science} \bibinfo{volume}{82}, \bibinfo{pages}{166--180}.
\bibitem[{Dutta and Husain(2009)}]{DuttaHu09}
\bibinfo{author}{Dutta, M.}, \bibinfo{author}{Husain, Z.},
  \bibinfo{year}{2009}.
\newblock \bibinfo{title}{An application of multicriteria decision making to
  built heritage. {T}he case of {C}alcutta}.
\newblock \bibinfo{journal}{Journal of Culural Heritage} \bibinfo{volume}{10},
  \bibinfo{pages}{237--243}.
\bibitem[{Fern\'andez et~al.(2017)Fern\'andez, Figueira, Navarro and
  Roy}]{FernandezEtAl17}
\bibinfo{author}{Fern\'andez, E.}, \bibinfo{author}{Figueira, J.R.},
  \bibinfo{author}{Navarro, J.}, \bibinfo{author}{Roy, B.},
  \bibinfo{year}{2017}.
\newblock \bibinfo{title}{{\sc{Electre Tri-nB}}: A new multiple criteria
  ordinal classification method}.
\newblock \bibinfo{journal}{European Journal of Operational Research}
  \bibinfo{volume}{263}, \bibinfo{pages}{214--224}.
\bibitem[{Ferretti et~al.(2014)Ferretti, Bottero and Mondini}]{FerrettiEtAl14}
\bibinfo{author}{Ferretti, V.}, \bibinfo{author}{Bottero, M.},
  \bibinfo{author}{Mondini, G.}, \bibinfo{year}{2014}.
\newblock \bibinfo{title}{Decision making and cultural heritage: {A}n
  application of the multi-attribute value theory for the reuse of historical
  buildings}.
\newblock \bibinfo{journal}{Journal of Cultural Heritage} \bibinfo{volume}{15},
  \bibinfo{pages}{644--655}.
\bibitem[{Ferretti and Comino(2015)}]{FerrettiCo15}
\bibinfo{author}{Ferretti, V.}, \bibinfo{author}{Comino, E.},
  \bibinfo{year}{2015}.
\newblock \bibinfo{title}{An integrated framework to assess complex cultural
  and natural heritage systems with multi-attribute value theory}.
\newblock \bibinfo{journal}{Journal of Cultural Heritage} \bibinfo{volume}{16},
  \bibinfo{pages}{688--697}.
\bibitem[{Figueira et~al.(2009)Figueira, Greco and Roy}]{FigueiraEtAl09}
\bibinfo{author}{Figueira, J.R.}, \bibinfo{author}{Greco, S.},
  \bibinfo{author}{Roy, B.}, \bibinfo{year}{2009}.
\newblock \bibinfo{title}{{\sc{Electre}} methods with interaction between
  criteria: {A}n extension of the concordance index}.
\newblock \bibinfo{journal}{European Journal of Operational Research}
  \bibinfo{volume}{199}, \bibinfo{pages}{478--495}.
\bibitem[{Figueira et~al.(2013)Figueira, Greco, Roy and
  S{\l}owi\'nski}]{FigueiraEtAl13}
\bibinfo{author}{Figueira, J.R.}, \bibinfo{author}{Greco, S.},
  \bibinfo{author}{Roy, B.}, \bibinfo{author}{S{\l}owi\'nski, R.},
  \bibinfo{year}{2013}.
\newblock \bibinfo{title}{An overview of {E}{\sc{lectre}} methods and their
  recent extensions.}
\newblock \bibinfo{journal}{Journal of Multi-Criteria Decision Analysis}
  \bibinfo{volume}{20}, \bibinfo{pages}{61--85}.
\bibitem[{Figueira et~al.(2016)Figueira, Mousseau and Roy}]{FigueiraEtAl16}
\bibinfo{author}{Figueira, J.R.}, \bibinfo{author}{Mousseau, V.},
  \bibinfo{author}{Roy, B.}, \bibinfo{year}{2016}.
\newblock \bibinfo{title}{{\sc{Electre}} methods}, in: \bibinfo{editor}{Greco,
  S.}, \bibinfo{editor}{Ehrgott, M.}, \bibinfo{editor}{Figueira, J.R.} (Eds.),
  \bibinfo{booktitle}{Multiple Criteria Decision Analysis - State of the Art
  Surveys}. \bibinfo{publisher}{Springer Science+Business Media, Inc.},
  \bibinfo{address}{New York, NY, USA}, pp. \bibinfo{pages}{155--185}.
\bibitem[{Figueira and Roy(2002)}]{FigueiraRo02}
\bibinfo{author}{Figueira, J.R.}, \bibinfo{author}{Roy, B.},
  \bibinfo{year}{2002}.
\newblock \bibinfo{title}{Determining the weights of criteria in the
  {E}{\sc{lectre}} type methods with a revised {S}imos' procedure}.
\newblock \bibinfo{journal}{European Journal of Operational Research}
  \bibinfo{volume}{139}, \bibinfo{pages}{317--326}.
\bibitem[{Garc{\'\i}a-Hern{\'a}ndez et~al.(2017)Garc{\'\i}a-Hern{\'a}ndez,
  de~la Calle-Vaquero and Yubero}]{garcia2017cultural}
\bibinfo{author}{Garc{\'\i}a-Hern{\'a}ndez, M.}, \bibinfo{author}{de~la
  Calle-Vaquero, M.}, \bibinfo{author}{Yubero, C.}, \bibinfo{year}{2017}.
\newblock \bibinfo{title}{Cultural heritage and urban tourism: Historic city
  centres under pressure}.
\newblock \bibinfo{journal}{Sustainability} \bibinfo{volume}{9},
  \bibinfo{pages}{1346}.
\bibitem[{Giove et~al.(2011)Giove, Rosato and Breil}]{GioveEtAl11}
\bibinfo{author}{Giove, S.}, \bibinfo{author}{Rosato, P.},
  \bibinfo{author}{Breil, M.}, \bibinfo{year}{2011}.
\newblock \bibinfo{title}{An application of multicriteria decision making to
  build heritage: {T}he redevelopment of {V}enice {A}rsenale}.
\newblock \bibinfo{journal}{Journal of Multi-Criteria Decision Analysis}
  \bibinfo{volume}{17}, \bibinfo{pages}{85--99}.
\bibitem[{Girard and De~Toro(2007)}]{GirardTo07}
\bibinfo{author}{Girard, L.F.}, \bibinfo{author}{De~Toro, P.},
  \bibinfo{year}{2007}.
\newblock \bibinfo{title}{Integrated spatial assessment: {A} multicriteria
  approach to sustainable development of cultural and environmental heritage in
  san marco dei cavoti, italy}.
\newblock \bibinfo{journal}{Central European Journal of Operations Research}
  \bibinfo{volume}{15}, \bibinfo{pages}{281--299}.
\bibitem[{Giuliani et~al.(2018)Giuliani, De~Falco, Landi, Bevilacqua, Santini
  and Pecori}]{GiulianiEtAl18}
\bibinfo{author}{Giuliani, F.}, \bibinfo{author}{De~Falco, A.},
  \bibinfo{author}{Landi, S.}, \bibinfo{author}{Bevilacqua, M.},
  \bibinfo{author}{Santini, L.}, \bibinfo{author}{Pecori, S.},
  \bibinfo{year}{2018}.
\newblock \bibinfo{title}{Reusing grain silos from the 1930s in {I}taly. {A}
  multi-criteria decision analysis for the case of {A}rezzo}.
\newblock \bibinfo{journal}{Journal of Cultural Heritage} \bibinfo{volume}{29},
  \bibinfo{pages}{145--159}.
\bibitem[{Gra{\v{z}}ulevi{\v{c}}i{\=u}t{\.e}(2006)}]{gravzulevivciute2006cultural}
\bibinfo{author}{Gra{\v{z}}ulevi{\v{c}}i{\=u}t{\.e}, I.}, \bibinfo{year}{2006}.
\newblock \bibinfo{title}{Cultural heritage in the context of sustainable
  development.}
\newblock \bibinfo{journal}{Environmental Research, Engineering \& Management}
  \bibinfo{volume}{37}, \bibinfo{pages}{74--79}.
\bibitem[{Greco et~al.(2016)Greco, Figueira and Ehrgott}]{greco2016multiple}
\bibinfo{author}{Greco, S.}, \bibinfo{author}{Figueira, J.},
  \bibinfo{author}{Ehrgott, M.}, \bibinfo{year}{2016}.
\newblock \bibinfo{title}{Multiple Criteria Decision Analysis: State of the Art
  Surveys}.
\newblock \bibinfo{publisher}{Springer}, \bibinfo{address}{Berlin}.
\bibitem[{Hamadouche et~al.(2014)Hamadouche, Mederbal, Kouri, Regagba, Fekir
  and Anteur}]{HamadoucheEtAl14}
\bibinfo{author}{Hamadouche, M.}, \bibinfo{author}{Mederbal, K.},
  \bibinfo{author}{Kouri, L.}, \bibinfo{author}{Regagba, Z.},
  \bibinfo{author}{Fekir, Y.}, \bibinfo{author}{Anteur, D.},
  \bibinfo{year}{2014}.
\newblock \bibinfo{title}{Gis-based multicriteria analysis: {A}n approach to
  select priority areas for preservation in the ahaggar national park,
  algeria}.
\newblock \bibinfo{journal}{Arabian Journal of Geosciences}
  \bibinfo{volume}{7}, \bibinfo{pages}{419--434}.
\bibitem[{Hansen et~al.(2015)Hansen, Frantzeskaki, McPhearson, Rall, Kabisch,
  Kaczorowska, Kain, Artmann and Pauleit}]{hansen2015uptake}
\bibinfo{author}{Hansen, R.}, \bibinfo{author}{Frantzeskaki, N.},
  \bibinfo{author}{McPhearson, T.}, \bibinfo{author}{Rall, E.},
  \bibinfo{author}{Kabisch, N.}, \bibinfo{author}{Kaczorowska, A.},
  \bibinfo{author}{Kain, J.H.}, \bibinfo{author}{Artmann, M.},
  \bibinfo{author}{Pauleit, S.}, \bibinfo{year}{2015}.
\newblock \bibinfo{title}{The uptake of the ecosystem services concept in
  planning discourses of european and american cities}.
\newblock \bibinfo{journal}{Ecosystem Services} \bibinfo{volume}{12},
  \bibinfo{pages}{228--246}.
\bibitem[{Hong and Chen(2017)}]{HongCh17}
\bibinfo{author}{Hong, Y.}, \bibinfo{author}{Chen, F.}, \bibinfo{year}{2017}.
\newblock \bibinfo{title}{Evaluating the adaptive reuse potential of buildings
  in conservation areas}.
\newblock \bibinfo{journal}{Facilities} \bibinfo{volume}{35(3-4)},
  \bibinfo{pages}{202--219}.
\bibitem[{Ishizaka and Nemery(2013)}]{ishizaka2013multi}
\bibinfo{author}{Ishizaka, A.}, \bibinfo{author}{Nemery, P.},
  \bibinfo{year}{2013}.
\newblock \bibinfo{title}{Multi-criteria decision analysis: methods and
  software}.
\newblock \bibinfo{publisher}{John Wiley \& Sons}.
\bibitem[{Kutut et~al.(2014)Kutut, Zavadskas and Lazauskas}]{KututEtAl14}
\bibinfo{author}{Kutut, V.}, \bibinfo{author}{Zavadskas},
  \bibinfo{author}{Lazauskas, M.}, \bibinfo{year}{2014}.
\newblock \bibinfo{title}{Assessment of priority alternatives for preservation
  of historic buildings using model based on {ARAS} and {AHP} methods}.
\newblock \bibinfo{journal}{Archives of Civil and Mechanic Engineering}
  \bibinfo{volume}{14}, \bibinfo{pages}{287--294}.
\bibitem[{Lolli et~al.(2017)Lolli, Ishizaka, Gamberini, Rimini, Balugani and
  Prandini}]{lolli2017}
\bibinfo{author}{Lolli, F.}, \bibinfo{author}{Ishizaka, A.},
  \bibinfo{author}{Gamberini, R.}, \bibinfo{author}{Rimini, B.},
  \bibinfo{author}{Balugani, E.}, \bibinfo{author}{Prandini, L.},
  \bibinfo{year}{2017}.
\newblock \bibinfo{title}{Requalifying public buildings and utilities using a
  group decision support system}.
\newblock \bibinfo{journal}{Journal of Cleaner Production}
  \bibinfo{volume}{164}, \bibinfo{pages}{1081--1092}.
\bibitem[{McGreevy(2017)}]{mcgreevy2017complexity}
\bibinfo{author}{McGreevy, M.P.}, \bibinfo{year}{2017}.
\newblock \bibinfo{title}{Complexity as the telos of postmodern planning and
  design: Designing better cities from the bottom-up}.
\newblock \bibinfo{journal}{Planning Theory} \bibinfo{volume}{17},
  \bibinfo{pages}{355–--374}.
\bibitem[{McKercher et~al.(2005)McKercher, Ho and Du~Cros}]{MckercherEtAl05}
\bibinfo{author}{McKercher, B.}, \bibinfo{author}{Ho, P.},
  \bibinfo{author}{Du~Cros, H.}, \bibinfo{year}{2005}.
\newblock \bibinfo{title}{Relationship between tourism and cultural heritage
  management: {E}vidence from hong kong}.
\newblock \bibinfo{journal}{Tourism Management} \bibinfo{volume}{26},
  \bibinfo{pages}{539--548}.
\bibitem[{Nestic\`o et~al.(2018)Nestic\`o, Morano and Sica}]{NesticoEtAl17}
\bibinfo{author}{Nestic\`o, A.}, \bibinfo{author}{Morano, P.},
  \bibinfo{author}{Sica, F.}, \bibinfo{year}{2018}.
\newblock \bibinfo{title}{A model to support the public administration
  decisions for the investments selection on historic buildings}.
\newblock \bibinfo{journal}{Journal of Cultural Heritage} \bibinfo{volume}{33},
  \bibinfo{pages}{201--207}.
\bibitem[{Newman(1999)}]{newman1999sustainability}
\bibinfo{author}{Newman, P.W.}, \bibinfo{year}{1999}.
\newblock \bibinfo{title}{Sustainability and cities: extending the metabolism
  model}.
\newblock \bibinfo{journal}{Landscape and Urban planning} \bibinfo{volume}{44},
  \bibinfo{pages}{219--226}.
\bibitem[{Oppio et~al.(2015)Oppio, Bottero, Ferretti, Fratesi and
  Ponzini}]{OppioEtAl15}
\bibinfo{author}{Oppio, A.}, \bibinfo{author}{Bottero, M.},
  \bibinfo{author}{Ferretti, V.}, \bibinfo{author}{Fratesi, U.},
  \bibinfo{author}{Ponzini, D.}, \bibinfo{year}{2015}.
\newblock \bibinfo{title}{Giving space to multicriteria analysis for complex
  cultural heritage systems: {T}he case of the castles in {V}alle {D}'{A}osta
  region, {I}taly}.
\newblock \bibinfo{journal}{Journal of Cultural Heritage} \bibinfo{volume}{16},
  \bibinfo{pages}{779--789}.
\bibitem[{Piano-Napoli(2011)}]{comune2011}
\bibinfo{author}{Piano-Napoli}, \bibinfo{year}{2011}.
\newblock \bibinfo{title}{Piano di gestione del centro storico di Napoli sito
  {UNESCO}}.
\newblock \bibinfo{type}{Technical Report}. \bibinfo{address}{Comune di Napoli,
  Napoli, Italia}.
\bibitem[{Rees and Wackernagel(1996)}]{rees1996urban}
\bibinfo{author}{Rees, W.}, \bibinfo{author}{Wackernagel, M.},
  \bibinfo{year}{1996}.
\newblock \bibinfo{title}{Urban ecological footprints: why cities cannot be
  sustainable—and why they are a key to sustainability}.
\newblock \bibinfo{journal}{Environmental impact assessment review}
  \bibinfo{volume}{16}, \bibinfo{pages}{223--248}.
\bibitem[{Roy(1996)}]{Roy96}
\bibinfo{author}{Roy, B.}, \bibinfo{year}{1996}.
\newblock \bibinfo{title}{Multicriteria Methodology for Decision Aiding}.
\newblock \bibinfo{publisher}{Kluwer Academic Publishers},
  \bibinfo{address}{Dordrecht, The Netherlands}.
\bibitem[{Roy(1998)}]{roy1998}
\bibinfo{author}{Roy, B.}, \bibinfo{year}{1998}.
\newblock \bibinfo{title}{A missing link in or-da: Robustness analysis}.
\newblock \bibinfo{journal}{Foundations of Computing and Decision Sciences}
  \bibinfo{volume}{23}, \bibinfo{pages}{141--160}.
\bibitem[{Roy(2010a)}]{roy2010robustness}
\bibinfo{author}{Roy, B.}, \bibinfo{year}{2010}a.
\newblock \bibinfo{title}{Robustness in operational research and decision
  aiding: A multi-faceted issue}.
\newblock \bibinfo{journal}{European Journal of Operational Research}
  \bibinfo{volume}{200}, \bibinfo{pages}{629--638}.
\bibitem[{Roy(2010b)}]{roy2010two}
\bibinfo{author}{Roy, B.}, \bibinfo{year}{2010}b.
\newblock \bibinfo{title}{Two conceptions of decision aiding}.
\newblock \bibinfo{journal}{International Journal of Multicriteria Decision
  Making} \bibinfo{volume}{1}, \bibinfo{pages}{74--79}.
\bibitem[{Roy and Bouyssou(1993)}]{RoyBo93}
\bibinfo{author}{Roy, B.}, \bibinfo{author}{Bouyssou, D.},
  \bibinfo{year}{1993}.
\newblock \bibinfo{title}{Aide Multicritère \`a la D\'ecision~: M\'ethodes et
  Cas}.
\newblock \bibinfo{publisher}{Economica}, \bibinfo{address}{Paris, France}.
\bibitem[{Roy et~al.(2014)Roy, Figueira and Almeida-Dias}]{RoyEtAl14}
\bibinfo{author}{Roy, B.}, \bibinfo{author}{Figueira, J.R.},
  \bibinfo{author}{Almeida-Dias, J.}, \bibinfo{year}{2014}.
\newblock \bibinfo{title}{Discriminating thresholds as a tool to cope with
  imperfect knowledge in multiple criteria decision aiding: {T}heoretical
  results and practical issues}.
\newblock \bibinfo{journal}{{\sc{Omega}}, The International Journal of
  Management Science} \bibinfo{volume}{43}, \bibinfo{pages}{9--20}.
\bibitem[{Roy and Vincke(1984)}]{RoyVi84}
\bibinfo{author}{Roy, B.}, \bibinfo{author}{Vincke, P.}, \bibinfo{year}{1984}.
\newblock \bibinfo{title}{Relational systems of preference with one or more
  pseudo-criteria: {S}ome new concepts and results}.
\newblock \bibinfo{journal}{Management Science} \bibinfo{volume}{30},
  \bibinfo{pages}{1323--1335}.
\bibitem[{Salo et~al.(2011)Salo, Keisler and Morton}]{SaloEtAl11}
\bibinfo{author}{Salo, A.}, \bibinfo{author}{Keisler, J.},
  \bibinfo{author}{Morton, A.}, \bibinfo{year}{2011}.
\newblock \bibinfo{title}{An invitation to portfolio decision analysis}, in:
  \bibinfo{editor}{Salo, A.}, \bibinfo{editor}{Keisler, J.},
  \bibinfo{editor}{Morton, A.} (Eds.), \bibinfo{booktitle}{Portfolio Decision
  Analysis}. \bibinfo{publisher}{Springer Science + Business Media, Inc.}, pp.
  \bibinfo{pages}{3--27}.
\bibitem[{Tarrag{\"u}el et~al.(2012)Tarrag{\"u}el, Krol and
  Van~Westen}]{TarraguelEtAl12}
\bibinfo{author}{Tarrag{\"u}el, A.}, \bibinfo{author}{Krol, B.},
  \bibinfo{author}{Van~Westen, C.}, \bibinfo{year}{2012}.
\newblock \bibinfo{title}{Analysing the possible impact of landslides and
  avalanches on cultural heritage in upper {S}vaneti, {G}eorgia}.
\newblock \bibinfo{journal}{Journal of Cultural Heritage} \bibinfo{volume}{13},
  \bibinfo{pages}{453--461}.
\bibitem[{Thabrew et~al.(2009)Thabrew, Wiek and
  Ries}]{thabrew2009environmental}
\bibinfo{author}{Thabrew, L.}, \bibinfo{author}{Wiek, A.},
  \bibinfo{author}{Ries, R.}, \bibinfo{year}{2009}.
\newblock \bibinfo{title}{Environmental decision making in multi-stakeholder
  contexts: applicability of life cycle thinking in development planning and
  implementation}.
\newblock \bibinfo{journal}{Journal of Cleaner Production}
  \bibinfo{volume}{17}, \bibinfo{pages}{67--76}.
\bibitem[{Tuan and Navrud(2008)}]{TuanNa08}
\bibinfo{author}{Tuan, T.}, \bibinfo{author}{Navrud, S.}, \bibinfo{year}{2008}.
\newblock \bibinfo{title}{Capturing the benefits of preserving cultural
  heritage}.
\newblock \bibinfo{journal}{Journal of Cultural Heritage} \bibinfo{volume}{9},
  \bibinfo{pages}{326--337}.
\bibitem[{Vasile et~al.(2015)Vasile, Surugiu and Stroe}]{VasileEtAl15}
\bibinfo{author}{Vasile, V.}, \bibinfo{author}{Surugiu, M.R.},
  \bibinfo{author}{Stroe, A.}, \bibinfo{year}{2015}.
\newblock \bibinfo{title}{Innovative valuing of the cultural heritage assets.
  {E}conomic implication on local employability, small entrepreneurship
  development and social inclusion}.
\newblock \bibinfo{journal}{Procedia-Social and Behavioral Sciences}
  \bibinfo{volume}{188}, \bibinfo{pages}{16--26}.
\bibitem[{Veldpaus et~al.(2013)Veldpaus, Pereira and
  Colenbrander}]{VeldpausEtAl13}
\bibinfo{author}{Veldpaus, L.}, \bibinfo{author}{Pereira, A.},
  \bibinfo{author}{Colenbrander, B.}, \bibinfo{year}{2013}.
\newblock \bibinfo{title}{Urban heritage: {P}utting the past into the future}.
\newblock \bibinfo{journal}{The Historic Environment: Policy \& Practice}
  \bibinfo{volume}{4}, \bibinfo{pages}{3--18}.
\bibitem[{Wang and Zeng(2010)}]{WangZe10}
\bibinfo{author}{Wang, H.J.}, \bibinfo{author}{Zeng, Z.T.},
  \bibinfo{year}{2010}.
\newblock \bibinfo{title}{A multi-objective decision-making process for reuse
  selection of historic buildings}.
\newblock \bibinfo{journal}{Expert Systems with Applications}
  \bibinfo{volume}{37}, \bibinfo{pages}{1241--1249}.
\bibitem[{Waterton(2015)}]{Waterton15}
\bibinfo{author}{Waterton, E.}, \bibinfo{year}{2015}.
\newblock \bibinfo{title}{Heritage and Community Engagement}.
  \bibinfo{publisher}{Springer Science+Business, Inc.}, \bibinfo{address}{New
  York, NY, USA}.
\newblock pp. \bibinfo{pages}{53--67}.
\bibitem[{Yung and Chan(2013)}]{YungCh13}
\bibinfo{author}{Yung, E.}, \bibinfo{author}{Chan, E.}, \bibinfo{year}{2013}.
\newblock \bibinfo{title}{Evaluation for the conservation of historic
  buildings: {D}ifferences between the laymen, professionals and policy
  makers}.
\newblock \bibinfo{journal}{Facilites} \bibinfo{volume}{31(11-12)},
  \bibinfo{pages}{542--564}.

\end{thebibliography}

\end{document}